\documentclass{aa}  

\usepackage{graphicx}
\usepackage{txfonts}
\usepackage{lipsum}
\usepackage{subcaption}         
\usepackage{lscape}             
\usepackage{placeins}      
                                
\usepackage{hyperref}

\usepackage[table]{xcolor}
\usepackage{tablefootnote}

\makeatletter
\def\linenumbers{%
  \@latex@error{Line numbers are disabled for this document}\@ehd
}
\def\@linenumbers{%
  \@latex@error{Line numbers are disabled for this document}\@ehd
}
\let\@linenumbers\relax
\let\linenumbers\relax
\makeatother

\begin{document}

\newcommand{\psj}{PSJ}
\newcommand{\dlink}{d_\mathrm{link}}
\newcommand{\phieff}{\phi_\mathrm{eff}}
\newcommand{\kcm}{\ \mathrm{kg/m}^{3}}

\titlerunning{Binary asteroid satellites born from sub-escape-velocity moonlet mergers}
\title{The diverse shapes of binary asteroid satellites born from sub-escape-velocity moonlet mergers}

\author{John Wimarsson\inst{1}\thanks{E-mail: \href{mailto:john.wimarsson@unibe.ch}{john.wimarsson@unibe.ch}}
\and Fabio Ferrari\inst{2,1}
\and Martin Jutzi\inst{1}}

\institute{Space Research \& Planetary Sciences, Physics Institute, University of Bern,
Gesellschaftsstrasse 6, 3012, Bern, Switzerland
\and Department of Aerospace Science and Technology, Politecnico di Milano,
20156, Milano, Italy}

\date{Received June 12, 2025 / Accepted October 9, 2025}

 \abstract 
   {Recent direct observations of atypically shaped rubble-pile satellites of sub-km asteroids in form of the spherically oblate Dimorphos and bilobate Selam challenge classical binary asteroid formation theories, which only explain the predominantly elongated population. This study further explores a rubble-pile satellite formation scenario for binary asteroid systems involving debris disks by investigating how mergers between moonlets with impact velocities below the mutual escape speed (sub-escape-velocity mergers) and tidal disruptions can create atypically shaped moons. We simulated sub-escape-velocity mergers between moonlets and studied the resulting structural evolution of the formed moon in a tidal environment using the polyhedral discrete elements method $N$-body code \texttt{GRAINS}. Firstly, we find that the shapes of rubble-pile moons formed by mergers in this regime are highly dependent on the shape and initial orientation of the involved moonlets. This can be explained by the moonlets largely retaining their individual structures during the impact. Secondly, we observe that mass-loss via tidal disruption for a bilobate object occurs in discrete regimes of distance to the primary. Closer to the primary, the innermost lobe is completely stripped off, while only a small piece of it is lost further out. Due to moonlets largely retaining their shape after undergoing a sub-escape-velocity merger, it is necessary to account for their non-sphericity to accurately model satellite formation in circumasteroidal debris disks. Moreover, the reshaping of merged objects via tidal disruption and distortion can produce oblate spheroid moons such as Dimorphos and highly elongated bilobate satellites with distinct necks such as Selam.}


\maketitle

\defcitealias{Wimarsson_et_al_2024}{W24}
\defcitealias{Agrusa_et_al_2024}{Agrusa et al. 2024}
\defcitealias{Madeira_&_Charnoz_2024}{Madeira \& Charnoz 2024}

\section{Introduction} \label{section:intro}

The existence of binary asteroid systems with small primaries is largely explained by rapidly rotating rubble-pile asteroids undergoing structural failure and shedding mass, either via ejection of boulders from the equator \citep{Walsh_et_al_2008,Walsh_et_al_2012}, contact binary fission \citep{Jacobson_&_Scheeres_2011a} or ejection of a large chunk of mass \citep{Tardivel_et_al_2018}, creating a satellite. The rapid rotation is likely caused by gradual spin-up via the YORP (Yarkovsky-O'Keefe-Radzievskii-Paddack) effect, where anisotropic re-emission of absorbed sunlight generates a weak torque that can either spin the asteroid up or down over tens of thousands to millions of years, depending on its size, shape and distance from the Sun \citep{Rubincam_2000}. However, while these formation models explain the predominantly prolate (elongated) population of asteroid satellites \citep{Pravec_et_al_2016}, the recent observations of two atypically shaped moons in binary asteroid systems have proved difficult to explain. First, NASA's DART mission directly imaged the spherically oblate Dimorphos orbiting the asteroid Didymos \citep{Rivkin_et_al_2021,Daly_et_al_2023a,Chabot_et_al_2024} and soon thereafter the Lucy spacecraft observed what appears to be a bilobate, contact binary satellite now called Selam during a fly-by of the Dinkinesh system \citep{Levison_et_al_2021,Levison_et_al_2024}. Their shapes prove difficult to reproduce with a single-moonlet formation model, as tidal forces render objects formed at typical distances from their primary prolate \citep{Porco_et_al_2007,Tiscareno_et_al_2013}. Instead, models involving circumasteroidal debris disks caused by chaotic mass-shedding events provide an alternative path to explain atypically shaped asteroids \citep{Madeira_&_Charnoz_2024,Agrusa_et_al_2024,Wimarsson_et_al_2024}, as such a dynamically rich environment can lead to low-velocity mergers, tidal disruptions and subsequent mass-accretion that deform satellites from their initially prolate shapes. For a more in-depth introduction to binary asteroid formation, we refer to \citet{Agrusa_et_al_2024} and \citet{Wimarsson_et_al_2024}, in which the authors explored the formation of atypically shaped asteroid satellites using Discrete Element Methods (DEM). 

More specifically, performing simulations with a polyhedral $N$-body code \citet{Wimarsson_et_al_2024} found that disks with a width of more than two primary radii form multiple moonlets that dynamically interact to form oblate spheroids and bilobate objects. Analysing mergers occurring in these debris disk simulations, they found that the initial non-spherical shapes, orientation and mass difference between the moonlets highly affected the final shape. Crucially, they further identified that moonlet--moonlet mergers occurring in this regime have impact velocities, $v_\mathrm{impact}$, below the mutual escape velocity, $v_\mathrm{esc}$, leading to little initial deformation of the separate moonlets. In contrast, other studies investigating low-speed mergers between similar-sized rubble-pile structures, forming e.g.~moons around Saturn \citep{Leleu_et_al_2018}, bilobate cometary nuclei \citep{Jutzi_&_Asphaug_2015,Jutzi_&_Benz_2017}, contact binary asteroids \citep{Mahronic_et_al_2021}, or asteroid satellites \citep{Raducan_et_al_2025}, mainly consider velocities of the order of one or a few escape velocities. The very low velocities observed by \citet{Wimarsson_et_al_2024}, with the highest being $0.86v_\mathrm{esc}$, can be explained by moonlets being on similar orbits, with eccentricities kept low by gravitational interactions and dissipative contacts with surrounding debris in the disk, leading to small impact angles $\leq 21.5^\circ$. As the impact velocity for moonlets A and B is proportional to $\mu^{-1/3}$, where $\mu = (m_\mathrm{A}+m_\mathrm{B})/M_\mathrm{primary}$ is the mutual mass ratio relative to the primary \citep{Leleu_et_al_2018}, this leads to the observed soft collisions. Such low impact velocities, on the order of or below $v_\mathrm{esc}$, have also been observed between clumps formed in gravitationally unstable circumstellar disks during planet formation simulations \citep{Schib_et_al_2025a,Schib_et_al_2025b}. To our best knowledge, strictly sub-escape-velocity mergers between rubble piles remain an unexplored regime that could explain the presence of atypically shaped satellites in binary asteroid systems.

To fully understand the dynamical history of observed asteroid satellites, it is also crucial to take reshaping via tidal forces into account, given how susceptible rubble piles are to elongation and mass-shedding. When assessing the stability of an object subject to tidal forces, a common approximation is to treat the body like a fluid according to the analysis by \citet{Roche_1847}. The fluid Roche limit is the closest approach where such a body can stay structurally intact. However, due to non-negligible friction levels in rubble piles, they exhibit higher survivability and are less prone to reshaping. There have been a significant number of studies investigating tidal disruption and distortion of rubble piles using numerical \citep{Asphaug_&_Benz_1994,Asphaug_&_Benz_1996,Richardson_Bottke_&_Love_1998,Bottke_et_al_1999,Walsh_&_Richardson_2006} as well as analytical methods \citep{Holsapple_&_Michel_2006,Holsapple_&_Michel_2008}, laying the groundwork for the current understanding. A general consensus of this research is that outcomes from tidal encounters depend on the distance to the primary object, the material's angle of friction, particle size frequency distribution (SFD), cohesion and the object shape. More recent works have further identified that particle packing, shape and simulation resolution also have a non-negligible effect on the outcome of tidal disruption events of rubble piles \citep{Movshovitz_et_al_2012,Zhang_&_Lin_2020,Zhang_&_Michel_2020,Marohnic_et_al_2023}.

With these results in mind, it is clear that tidal evolution of rubble piles is a dynamically rich subject, given how sensitive the outcome is to the specific properties of a model. Moreover, a new dimension of complexity is added to the problem when involving non-homogeneous shapes such as bilobate structures. In this work, we make a first attempt to narrow down the formation of atypically shaped asteroid moons via sub-escape-velocity mergers and their subsequent structural evolution under the influence of tides with the use of numerical $N$-body simulations. Our formation scenarios are based on the results from the debris disk model of \citet{Wimarsson_et_al_2024}, henceforth referred to as \citetalias{Wimarsson_et_al_2024}. In Sect.~\ref{section:method}, we introduce the numerical model and simulation setup, while Sect.~\ref{section:results} contains the results from our investigation. Section~\ref{section:discussion} provides analysis and discussion of how our results relate to the different aspects of our model and ties to potential formation scenarios of Dimorphos and Selam. Finally, Sect.~\ref{section:conclusions} gives our conclusions and outlook for future studies.

\section{Method} \label{section:method}

To track deformation of the moons resulting from the mergers, while capturing key granular dynamics and long-lasting contacts between boulders, we chose to base our simulations on the gravitational $N$-body code \texttt{GRAINS} \citep{Ferrari_et_al_2017,Ferrari_et_al_2020}, which uses irregularly shaped particles. We now go through the numerical details, followed by the implementation of the mergers in simulations.

\subsection{Numerical setup}\label{section:method_numerics}

\texttt{GRAINS} is based on the libraries of \texttt{Chrono::Engine}, an open source \texttt{C++} physics engine \citep[\texttt{Chrono,}][]{Chrono2016} and can use non-spherical particles to study rubble-pile dynamics in space. The angular nature of the particles allows us to replicate complex physical effects of granular media such as particle--particle interlocking, particle spin orientation, off-centre collisions and polyhedral contacts. Several studies have explored the benefit of using non-spherical particles, concluding that it enhances the ability to capture key dynamical behaviours of rubble-pile aggregates \citep{Korycansky_&_Asphaug_2006,Korycansky_&_Asphaug_2009,Ferrari_&_Tanga_2020,Marohnic_et_al_2023}. \texttt{GRAINS} is a DEM code that can handle tens of thousands of particles due to GPU acceleration and has been used to study rubble-pile aggregation \citep{Ferrari_et_al_2017,Ferrari_&_Tanga_2020} and rotational equilibrium \citep{Ferrari_&_Tanga_2022}, the dynamical state of the Didymos binary system \citep{Agrusa_et_al_2022,Ferrari_et_al_2022} and the formation of binary asteroid systems post rotational failure \citepalias{Wimarsson_et_al_2024}. Furthermore, it has recently been verified as a tool to study tidal dissipation in binary asteroid systems \citep{Burnett_Fodde_&_Ferrari_2025}. To resolve physical contacts between particles, \texttt{Chrono} offers two different options: non-smooth, which is impulse-based and smooth, which is force-based. The former performs well for short-lasting, high strain-rate events, while the smooth-contact method (SMC) is better suited for longer-duration contacts and low-velocity encounters, making it optimal for the purposes of this study. Moreover, using \texttt{GRAINS} with SMC makes it behave comparably to other DEM $N$-body codes that resolve contacts with soft-sphere methods \citep{Sanchez_&_Scheeres_2011,Schwartz_et_al_2012}.

The numerical configuration for our simulations is practically identical to the one in \citetalias{Wimarsson_et_al_2024}. We set a static time step of 0.2 s, such that it is smaller than the characteristic time of the fundamental frequency of the spring-dashpot contact model \citep{Ferrari_et_al_2020} and used the default parallel integrator of \texttt{Chrono}, which is a Nesterov accelerated projected gradient descent method \citep{Chrono_integrators2017}. To resolve contacts with SMC, we employed the non-linear Hertzian version of the two-way normal-tangent spring–dashpot system. The resulting forces due to contacts between two bodies depend on a set of constants that define physical properties of the material used. These include static and sliding friction, stiffness, damping, adhesion and restitution, which have been kept at their default values determined from comparisons with laboratory studies of granular media \citep{Chrono2016}. Aggregates in our simulations are strictly kept together due to gravity and granular dynamical mechanisms such as interlocking, meaning the adhesion is kept at zero. Note that contacts with the primary body in the system have been omitted to speed up our simulations.

\subsection{Particle generation}

In \texttt{GRAINS}, each simulation is populated with randomly generated non-spherical particles created by filling a box with a pre-selected number of points, $N_\mathrm{vertices}$, that act as vertices for a convex hull. The size of each box is determined by drawing a random diameter from the following exponential SFD

\begin{equation}\label{equation:particle_sfd}
    f_{D_\mathrm{p}}(D_\mathrm{p},D_\mathrm{min},D_\mathrm{mean}) = e^{-(D_\mathrm{p}-D_\mathrm{min})\lambda_D},
\end{equation}
\noindent
where $\lambda_D = 1/(D_\mathrm{mean}-D_\mathrm{min})$. Due to the random nature of the vertex generation, some convex hulls will end up with fewer than $N_\mathrm{vertices}$. For the simulations from \citetalias{Wimarsson_et_al_2024}, they employed an SFD with values of $D_\mathrm{mean}=30$ m and $D_\mathrm{min}=5$ m to keep the particle number of the generated debris disk on the order of $10^3$-$10^4$. The mass of each particle is calculated based on its generated volume and a previously defined particle density, $\rho_\mathrm{p}$, which was set to 1190 kg m$^{-3}$ in \citetalias{Wimarsson_et_al_2024}. Throughout the remainder of this paper, we also make use of the property, $\rho_\mathrm{bulk}$, which is the bulk density of a given aggregate.

\subsection{Simulation setup}\label{section:method_simulation}

\subsubsection{General configuration}

\begin{table}
    \caption{Properties of the \texttt{Ryugu\_run4} merger from \citet{Wimarsson_et_al_2024} in terms of mass ratio, velocity, distance from the primary and impact angle.}
    \label{tab:W24_merger_data}
    \centering
    \begin{tabular}{l l l l}
        \hline\hline
         Mass ratio & $v_\mathrm{impact}$ [$v_\mathrm{esc}]$ & $r$ [$R_\mathrm{primary}$] & Impact angle [$\degr$]\\
         \hline
          0.56 & 0.84 & 2.43 & 21.5 \\
         \hline 
    \end{tabular}
\end{table}

\begin{table*}[h]
    \caption{Masses, number of particles and DEEVE principle axis ratios for the two moonlets prior to their mergers.}
    \label{tab:moonlet_initial_data}
    \centering
    \begin{tabular}{l l l l l l l l}
        \hline\hline
         Body & $m_\mathrm{moonlet}\ [M_\mathrm{primary}]$ & $N_\mathrm{p}$ & $a/b$ & $b/c$ & $a\ [R_\mathrm{primary}]$ & $\rho_\mathrm{p}$ [kg m$^{-3}$] & $\rho_\mathrm{bulk}$ [kg m$^{-3}$]\\
         \hline
         Moonlet A & 0.0100 & 1147 & 1.50 & 1.26 & 0.34 & 1190 & 626.8 \\
         Moonlet B & 0.0054 & 802 & 1.43 & 0.76 & 0.22 & 1190 & 654.9 \\
         \hline 
    \end{tabular}
\end{table*}

\begin{figure}
    \resizebox{\hsize}{!}{\includegraphics{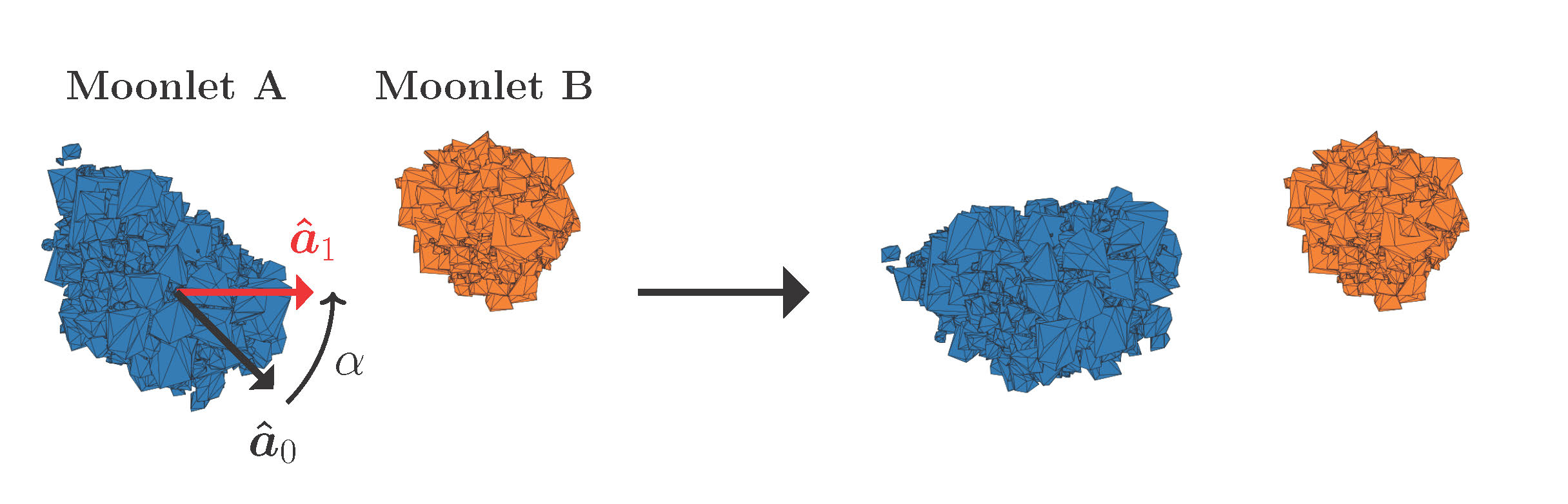}}
    \caption{Visualisation of how new initial conditions are generated by rotating the most massive moonlet counter-clockwise around its barycenter by an angle $\alpha$, here set to 45$^\circ$. On the left is the original configuration from the \citetalias{Wimarsson_et_al_2024} merger, with a vector $\boldsymbol{\hat{a}}_0$ parallel to the aggregate's semi-major axis. We then perform the rotation for each particle, ending up with a new aggregate with its semi-major axis parallel to the vector $\boldsymbol{\hat{a}}_1$.}
    \label{fig:moonlet_rotation_diagram} 
\end{figure}

The core part of our study revolved around altering the initial conditions of a specific merger and its subsequent dynamical evolution to better understand how pre-impact asymmetric, non-spherical shapes of the participating moonlets affect the shape of the final moon. The merger in question was observed during the later stages of a binary asteroid formation simulation based on the debris disk model in \citetalias{Wimarsson_et_al_2024}, namely \texttt{Ryugu\_run4}, which resulted in an elongated, bilobate satellite (see their Sect.~3.3.1 and 5.2). For the remainder of this paper, we refer to this event as the W24 merger and the dynamical details of the event are provided in Table~\ref{tab:W24_merger_data}. Isolating the two moonlets involved in the merger from the surrounding debris disk 0.5 h prior to their impact, using the friends-of-friends clustering method introduced in \citetalias{Wimarsson_et_al_2024}, we obtained a more general scenario where we could disregard the dynamical influence of the disk and any smaller aggregates present in the system. The configuration is shown in the left plot of Fig.~\ref{fig:moonlet_rotation_diagram} while the moonlet data can be found in Table~\ref{tab:moonlet_initial_data}. The most massive moonlet, situated closer to the primary, is henceforth known as moonlet A, while the less massive, more spherically oblate body is moonlet B. The principal axes of the moonlets, along with any reference to principal axes in the remainder of the paper, have been computed with the dynamically equivalent equal volume ellipsoid (DEEVE) method used in e.g.~\citet{Agrusa_et_al_2024} and \citetalias{Wimarsson_et_al_2024}. The axes are denoted $a,\,b$ and $c$ and follow $a>b$, with $c$ pointing out of the equatorial plane of the primary (for a full definition see Sect.~3.2 of \citetalias{Wimarsson_et_al_2024}). An oblate body has a principal axis ratio $a/b$ close to unity, with Dimorphos having the value $a/b=1.06\pm0.03$ \citep{Daly_et_al_2024}. To avoid confusion, we refer to the orbital semi-major axis as $a_\mathrm{orb}$ throughout the rest of this paper. 

To generate a set of initial configurations for the two moonlets that could help emphasise the effect of impact geometry for the post-merger shape of the moon, we imposed different rotations, $\alpha$, onto the bodies in the equatorial plane of the primary. Information on how rotations are handled in \texttt{GRAINS} can be found in Appendix~\ref{appendix:quaternions}. For a visual depiction of the rotation, see Fig.~\ref{fig:moonlet_rotation_diagram}. Noting that rotating the second aggregate had little to no effect on the final shape of the merged object, we focused only on altering the pre-impact state of moonlet A. Each component of the aggregate was rotated around its own $z$-axis, while also transforming its position, velocity and angular velocity relative to the barycentre of the moonlet. The result was a shift in orientation of the rubble pile while keeping the magnitude and direction of its angular and linear velocity vectors intact. All-in-all, we generated 24 different setups with counter-clockwise rotations $\alpha \in [0,360)^\circ$ using an increment of $15^\circ$, with $\alpha = 0^\circ$ being the fiducial configuration taken from \citetalias{Wimarsson_et_al_2024} seen on the left side of Fig.~\ref{fig:moonlet_rotation_diagram}.

\subsubsection{Higher resolution simulations}

\begin{table*}[h]
    \caption{Masses, number of particles and DEEVE principle axis ratios for the two moonlets used in the higher-resolution simulations, prior to their mergers.}
    \label{tab:moonlet_initial_data_highres}
    \centering
    \begin{tabular}{l l l l l l l l}
        \hline\hline
         Body & $m_\mathrm{moonlet}\ [M_\mathrm{primary}]$ & $N_\mathrm{p}$ & $a/b$ & $b/c$ & $a\ [R_\mathrm{primary}]$ & $\rho_\mathrm{p}$ [kg m$^{-3}$] & $\rho_\mathrm{bulk}$ [kg m$^{-3}$]\\
         \hline
         Moonlet A2 & 0.0100 & 23157 & 1.47 & 1.21 & 0.37 & 1134 & 626.8 \\
         Moonlet B2 & 0.0055 & 12572 & 1.31 & 0.84 & 0.24 & 1134 & 626.8 \\
         \hline 
    \end{tabular}
\end{table*}

As mentioned, the dynamics and structural stability of rubble piles are highly dependent on particle shapes and polyhedral contacts \citep{Korycansky_&_Asphaug_2006,Korycansky_&_Asphaug_2009,Movshovitz_et_al_2012,Ferrari_&_Tanga_2020,Marohnic_et_al_2023}, as well as SFD and boulder packing \citep{Zhang_et_al_2017,Zhang_et_al_2021,Zhang_&_Michel_2020,Raducan_et_al_2024b}. Hence, it is likely that the impact and tidal forces affecting the final moon will lead to different degrees of distortion when employing finer or coarser SFDs in our system. To elaborate, when considering non-spherical particles, if there is a uniform distribution of large and small boulders, the aggregate will be rigid as compared to a case where the distribution is dominated by smaller particles. This comes from the fact that the small particles can fill out voids surrounding larger boulders and smooth out contacts between them, making the aggregate less rigid. This mechanism has, e.g.~proven to facilitate propagation of seismic waves in granular material \citep{Marti_et_al_2024}. In turn, we set up simulations, using similar initial conditions but with aggregates of a steeper SFD with $D_\mathrm{mean}=20$ m and $D_\mathrm{mean}=3$ m, significantly increasing the particle number for each moonlet. The new SFD being more similar to observed SFDs on the surface of e.g.~Dimorphos and Didymos \citep{Barnouin_et_al_2024,Robin_et_al_2024,Pajola_et_al_2024} than the old, we chose to not treat our particles as porous collections of smaller sized boulders as in \citetalias{Wimarsson_et_al_2024} and opted for a higher material density of 3000 kg m$^{-3}$ (instead of 1190 kg m$^{-3}$) more similar to that of LL chondrites \citep{Flynn_et_al_2018}. 

This was done by filling a box with a large cloud of 80\,000 boulders with random positions and diameters, using the new SFD and material density. Letting the cloud gravitationally collapse into a clump, we could impose a three-dimensional graphical object model of each moonlet to create a higher-resolution moonlet with the correct shape. Each object model was created using the $\alpha$-wrap library of \texttt{CGAL} \citep[the computational geometry algorithms library,][]{cgal:alpha_wrap_3}, essentially wrapping a thin ``sheet'' around the moonlet, recursively subdividing its edges around the boulders until the algorithm has reached a target minimum length. Using built-in ray tracing capabilities of \texttt{Chrono}, we iterated through each body in the clump, sending out straight rays in the negative and positive directions of each axis from the centre-of-mass of the body. If all rays hit the alpha wrap mesh, the boulder was considered part of the new moonlet. As populating the shape with boulders from a finer SFD will lead to a different packing, we had to ensure the final configuration would have similar properties to the target moonlet in terms of bulk density and mass. This was achieved by exploiting the non-dimensionality of \texttt{GRAINS}. The resulting properties of each new aggregate, referred to as moonlets A2 and B2, can be found in Table~\ref{tab:moonlet_initial_data_highres}. We point out that both aggregates had to be given the same bulk and material densities due to limitations of the implementation in \texttt{GRAINS}. For each simulation, we integrated the initial 0.05 h with a small time step of 0.01 s to let the body properly settle, avoiding any divergent solutions for contact computation. To account for the change in particle SFD, the rest of the simulation was performed with a time step of 0.05 s.

\section{Results} \label{section:results}

\subsection{Sub-escape-velocity mergers}

\begin{figure}
    \resizebox{\hsize}{!}{\includegraphics{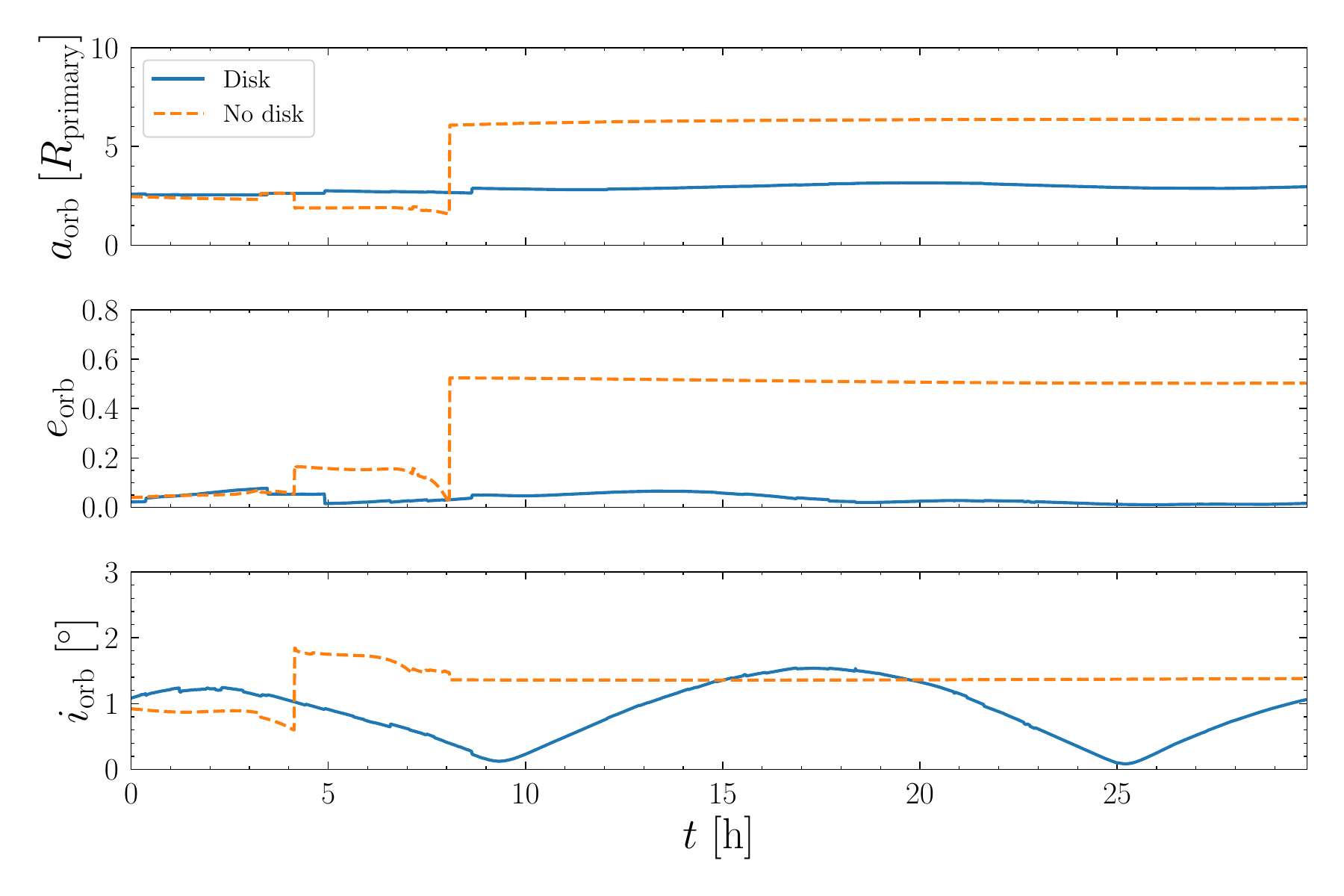}}
    \caption{Evolution of semi-major axis (top), eccentricity (middle) and inclination (bottom) for the moons formed in the \citetalias{Wimarsson_et_al_2024} merger with (solid blue) and without a disk present (dashed orange).}
    \label{fig:disk_vs_no_disk}
\end{figure}

\begin{figure*}
    \resizebox{\hsize}{!}{\includegraphics{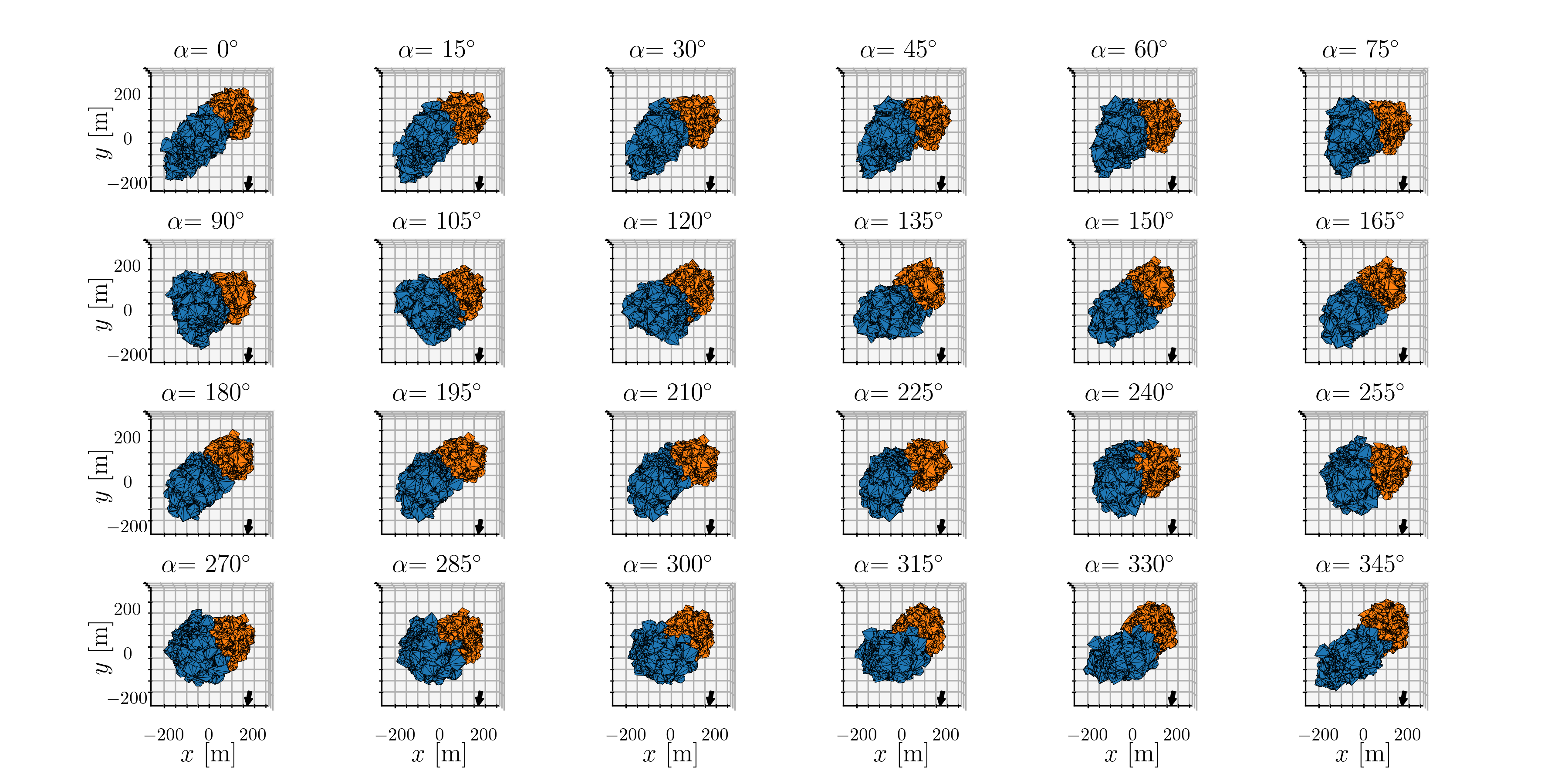}}
    \caption{Shapes and orientations of each formed moon after 3 h of simulation time after a sub-escape-velocity merger between the two moonlets in Table~\ref{tab:moonlet_initial_data}. The value $\alpha$, represents the counter-clockwise rotation relative to the initial position of moonlet A in Fig.~\ref{fig:moonlet_rotation_diagram} around its barycenter. The particles originally belonging to moonlet A have been coloured in blue, while the corresponding particles from moonlet B are orange. The black arrow in each plot is pointing in the direction of the primary.}
    \label{fig:merged_moon_3h}
\end{figure*}

Letting each initial configuration dynamically evolve for several orbits, it became clear that the removal of the disk significantly altered the resulting post-merger orbits for the newly formed moons, destabilising them. In Fig.~\ref{fig:disk_vs_no_disk}, we have plotted the post-merger evolution of the main orbital elements for the original \citetalias{Wimarsson_et_al_2024} case with a disk, as well as the scenario where the disk has been removed. The small initial offset in semi-major axis, eccentricity and inclination can be explained by the simulations being started 0.5 h before the merger, while only the evolution of the largest remnant after the merger is shown. As the two moons evolved, the dynamical effect of the remaining disk, which at this point had a mass of $0.031M_\mathrm{primary}$ (including $0.023M_\mathrm{primary}$ in other moonlets), became substantial. The difference is clearly seen when comparing the change in eccentricity over time. The two small declines seen for the disk case at the 3.5 h and near the 5 h mark were caused by two minor tidal disruption events, where the body only lost a few per cent of its mass in total. Any further excitation of the orbit was then dampened by the disk. In contrast, the no-disk moon suffered a catastrophic tidal disruption between 4.5 and 8 h, losing more than half of its mass as the bilobate structure got pulled apart. This accounts for the corresponding sharp increase in eccentricity and semi-major axis as only moonlet B remained, although on an eccentric orbit (we recommend watching the two complementary movies of this process\footnote{\texttt{alpha0deg\_merger\_lowres.avi} shows the merger in an inertial frame. For the following tidal disruption of the no-disk scenario, see \texttt{alpha0\_post\_merger\_lowres\_r0.avi}.}). The final properties of the moon were $m_\mathrm{moon} = 0.35M_0$, where $M_0$ is the post-merger mass of the progenitor, $a/b = 1.48$, $b/c = 0.78$ and $a=0.23R_\mathrm{primary}$. The merged moons for the remaining $\alpha$ values also suffer significant mass-loss in a majority of the cases, leaving oblate or prolate remnants behind (see Fig.~\ref{fig:merged_moon_48h_1p0} and Table~\ref{tab:moon_properties_48h_1p0}). All-in-all, this result indicates that bilobate moons formed in a disk have a higher survivability and that they can form closer to the primary than in the no-disk scenario.

To address the discrepancy in orbital behaviour, while still being able to investigate the stability and shapes of moons formed via sub-escape-velocity mergers, we decided to divide our study into two stages. First, we carried out each merger according to the 24 different initial conditions obtained via rotating moonlet A, starting each simulation 0.5 h before impact. Letting each formed moon settle after the merger, we paused the simulation after a total time of 3 h. Second, we pushed each moon further out from the primary, giving it a velocity corresponding to a circular orbit, leading to a new position of $\boldsymbol{r} = k\boldsymbol{r}_0$, where $k$ is a factor larger than 1 and $\boldsymbol{r}_0$ is the position relative the primary barycenter after 3 h. We cannot conclude that the mergers would lead to the same configurations when taking place further out from the primary without running new, self-consistent simulations. However, due to the chaotic nature of the debris disk formation scenario, along with the observed range of impact velocities recorded in \citetalias{Wimarsson_et_al_2024}, it is plausible that the shapes formed in our mergers could also form at these new distances $k\boldsymbol{r}_0$. Even so, the results from the second stage of the study have to be interpreted with the removal of the disk in mind. To clarify, it only teaches us about the structural and orbital stability of a given object in isolation, as the presence of the disk would alter the dynamics of the problem and lead to further deformation via subsequent accretion, potentially homogenising the surface and shape as seen in \citet{Agrusa_et_al_2024} and \citetalias{Wimarsson_et_al_2024}. That being said, in \citetalias{Wimarsson_et_al_2024}, the final shapes of the moons were largely determined by mergers, while subsequent accretion of single particles and small clusters had little effect on the final $a/b$ ratio in comparison.

\begin{figure}
    \centering
    \resizebox{0.5\hsize}{!}{\includegraphics{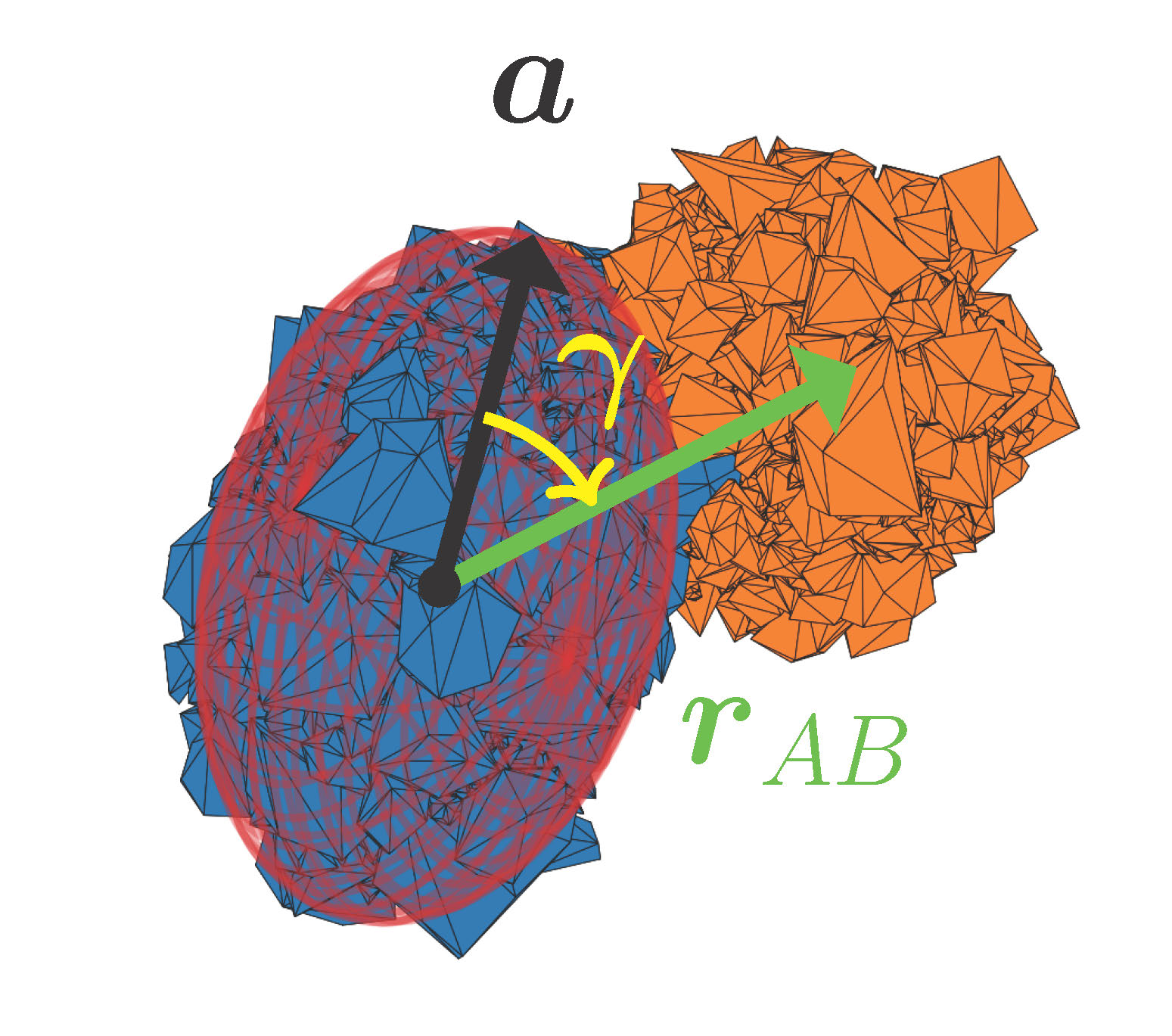}}
    \caption{The angle, $\gamma$, between the longest principal axis of the DEEVE of moonlet A (blue), $\boldsymbol{a}$, and the vector $\boldsymbol{r}_{AB}$, which is the positional vector between the barycenters of moonlet A and moonlet B (orange).}
    \label{fig:2DEEVE_classification_diagram}
\end{figure}

The initial shapes post-merger at the end of the first stage are shown in Fig.~\ref{fig:merged_moon_3h}. For each case, we have coloured the particles depending on which moonlet they belonged to before impact. All figures depicting moon shapes in this work have been rendered in the inertial equatorial plane of the primary. For impact velocities higher than $v_\mathrm{esc}$, we would expect the initial shapes of the bodies to have little effect on the final configuration of particles as mergers in this regime lead to significant deformation \citep{Leleu_et_al_2018,Raducan_et_al_2025}. However, for the case of sub-escape-velocity, the deformation due to the impact is small, and we observe drastic differences in post-merger shape with just $15^\circ$ rotations of the most elongated body, pre-merger, emphasising the influence of merger geometry. Among the resulting shapes, we can identify different types of bilobate structures, and we classify these using three different categories. The categories are quantified by $\gamma\in (-90,90]^\circ$, which represents the angle between the longest principal axis of the most massive lobe DEEVE fit (see Appendix \ref{appendix:DEEVE_fits}), and the vector between the two barycenters of the lobes, $\boldsymbol{r}_{AB}$, as shown in Fig.~\ref{fig:2DEEVE_classification_diagram}. The three categories are defined as follows: 

\begin{itemize}
    \item long-axis bilobate (LAB), where $|\gamma| < 10^\circ$, $\alpha=$ 0, 165, 180, 345$^\circ$;
    \item off-axis bilobate (OAB), where $10^\circ \leq |\gamma| < 45^\circ$, $\alpha =$ 15, 30, 45, 150, 195, 210, 225, 255, 330$^\circ$;
    \item short-axis bilobate (SAB), where $|\gamma| \geq 45^\circ$, $\alpha=$ 60, 75, 90, 105, 120, 135, 240, 255, 270, 285, 300, 315$^\circ$.
\end{itemize}

By this classification, Selam is potentially a LAB type \citep{Levison_et_al_2024}, while Itokawa \citep{Fujiwara_et_al_2006} is an OAB binary. The seemingly bilobate nature of Apophis \citep{Brozovic_et_al_2018,Brozovic_et_al_2022} would make it an SAB. We acknowledge that this methodology can only serve as a first attempt to classify these objects, as it fails to quantify how distinctly bilobate a merged object is. In turn, we aim to improve upon this classification system in further work. 

\begin{figure*}
    \resizebox{\hsize}{!}{\includegraphics{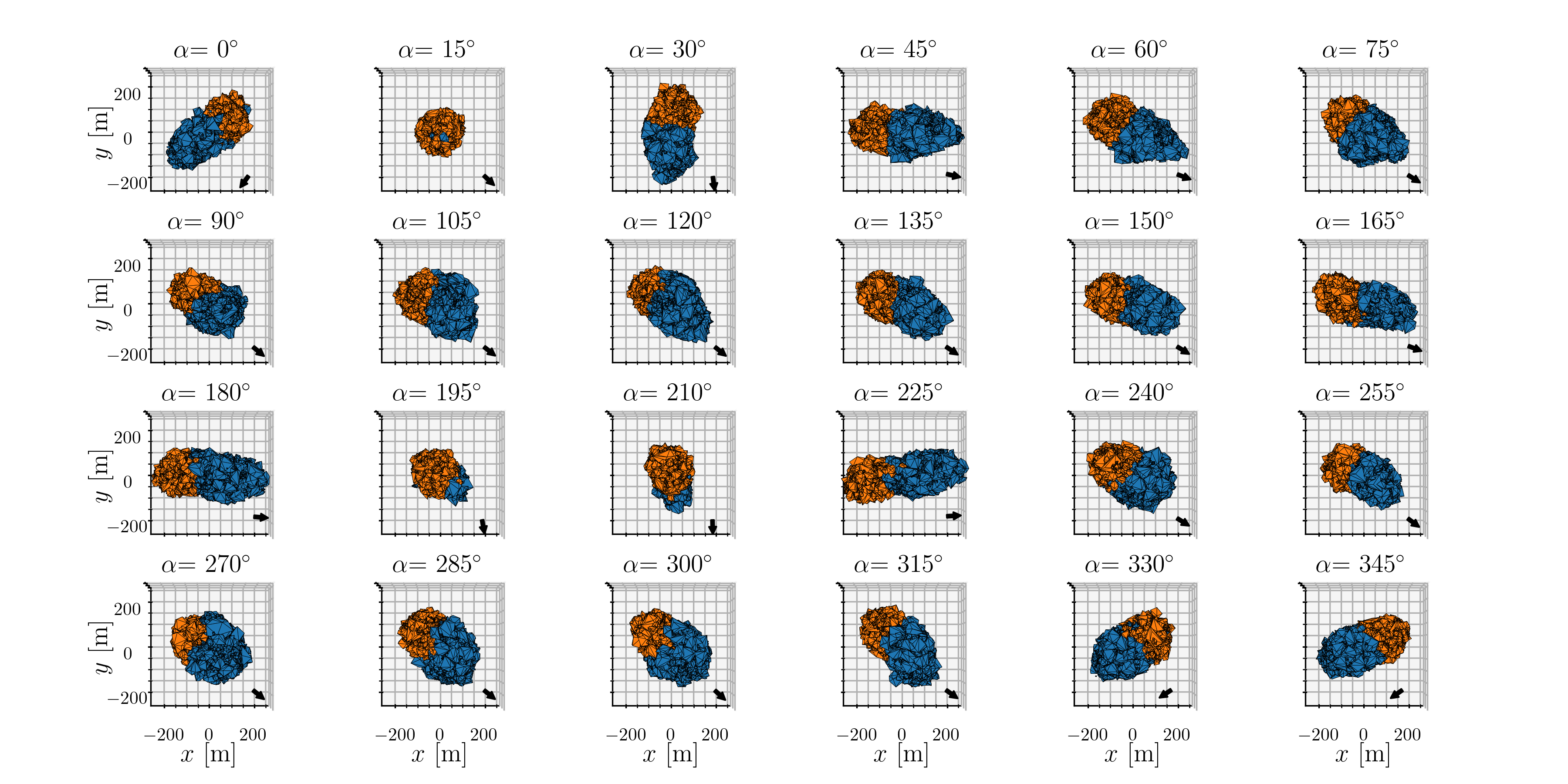}}
    \caption{Same as Fig.~\ref{fig:merged_moon_3h} but after putting each body on a circular orbit with a semi-major axis of $1.1r_0$ and evolving the system for an additional 45 h (48 h in total), where $r_0$ is the distance from the primary barycenter at 3 h.}
    \label{fig:merged_moon_48h_1p1}
\end{figure*}

\subsection{Survivability of formed objects}

As we could say little about the stability of each object at this point, we proceeded to evolve the system in the second stage. Running test simulations without the disk, placing the $\alpha=0^\circ$ moon at different distances from the primary (further explored in Section \ref{section:discussion_tidal_study}), we found that the orbital and structural evolution of the object was highly similar to the disk-case when $r \sim 1.1r_0$. Hence, for the first set of simulations, we let $r_1=1.1r_0=2.673 R_\mathrm{primary}$ for each $\alpha$ case and propagated the system for an additional 45 h. The resulting shapes after 48 h of total simulation are shown in Fig.~\ref{fig:merged_moon_48h_1p1}. The physical and orbital properties of each final moon can be found in Table~\ref{tab:moon_properties_48h_1p1}, along with reference principal axis values for the shapes at the onset of this second stage of simulation from Fig.~\ref{fig:merged_moon_3h}, subscripted by 0. We have attempted to classify the final shapes using our bilobate categories and $\gamma$ prescription, along with the two additional types: oblate spheroid (OS), i.e.~Dimorphos-like, and prolate spheroid (PS). The latter signifies a more compact shape where the bilobate structure of an elongated object can no longer be detected. Comparing the figures and properties of each moon, we observe how, throughout three orbits, the tidal influence from the primary has led to mass-shedding and disruption in some cases, as well as homogenisation of some bilobate shapes, smoothing out their surfaces. For the cases of $\alpha = $ 15, 195, 210$^\circ$, the disruption was catastrophic, resulting in mass-loss. Studying their shapes in Fig.~\ref{fig:merged_moon_48h_1p1}, we observe that most of moonlet A has been stripped off, along with some particles belonging to moonlet B, leaving behind an OS Dimorphos-like structure. While its DEEVE principal axes indicate $a/b=1.33$, we recall that this method for evaluating physical shapes can sometimes overestimate the $a/b$ value \citep{Agrusa_et_al_2024} and only serves as an estimate for the true physical extents. Formation of OS moons due to catastrophic tidal disruption was theorised in \citetalias{Wimarsson_et_al_2024}, but not shown until now. While this particular scenario leads to the object being put on an eccentric orbit, given the results from Fig.~\ref{fig:disk_vs_no_disk}, it is plausible that the presence of a disk would be enough to dampen the orbit and circularise it. Furthermore, it also becomes evident that LABs and OABs are more susceptible to disruption and mass loss, given that all SABs stay intact. Instead, the tidal forces from the primary act to ``straighten'' and smooth out the shapes of the short-axis merger objects, e.g.~$\alpha = 120^\circ$, which is transformed from an SAB to a more homogeneous PS without any real indications of being produced by a merger. We can get a sense of how the shape changes over time by comparing the initial and final values for $a/b$ and macroscopic porosity, estimated by evaluating the total volume of each aggregate with an $\alpha$-shape scheme\footnote{An $\alpha$-shape is essentially a shape generated by ``rolling a ball'' over a set of points, in our case the vertices of each convex hull representing our grains. The value of $\alpha_\mathrm{3D}$ is the squared radius of the ball and determines the resolution of the resulting mesh. For values $\alpha_\mathrm{3D}\to\infty$, the $\alpha$-shape is a convex hull \citep{Edelsbrunner_&_Mucke_1994}.} based on the \texttt{CGAL::Alpha\_shape\_3} library \citep{cgal:alpha_shapes_3}. This module possesses the ability to compute an optimal $\alpha_\mathrm{3D}$ value, which is the smallest value for which the algorithm will produce a ``watertight'' mesh (for a more in-depth explanation, see Sect.~2.3.1 of \citetalias{Wimarsson_et_al_2024}). In turn, the porosity calculated using this method is the minimum macroscopic porosity possible for our aggregates. While the $a/b$ values stay similar, except for extreme cases of either catastrophic disruption ($\alpha =$ 15, 195, 210$^\circ$) or major reshaping ($\alpha = 225^\circ$), the porosity is generally significantly lower for each case, emphasising how the tidal forces serve to, with a few exceptions, homogenise the aggregates and make them less elongated at these distances.

To further improve our understanding of how the distance to the primary affected the final shapes of the merged moons, we also performed the same set of rotationally altered simulations for the case of $r_2=1.2r_0$. The corresponding shapes after 48 h of total simulation are shown in Fig.~\ref{fig:merged_moon_48h_1p2} and their physical, geometrical and orbital properties can be found in Table~\ref{tab:moon_properties_48h_1p2}. Notably, the moons undergo no mass-shedding at these distances and mainly reshape due to homogenisation and spin, as a majority of them end up in super-synchronous rotational states (i.e.~their rotational periods are shorter than their orbital periods). Studying the change in $a/b$ values between the 3 h and 48 h mark, we find that the $a/b$-values consistently decrease for each $\alpha$ angle while the changes in porosity remain similar to that of the $1.1r_0$ cases. When further shifting the post-merger distances to 1.3$r_0$, tides become even less influential, leading to additional reductions in $a/b$.

\subsection{Higher resolution shapes}\label{section:results_highres}

\begin{figure}
    \resizebox{\hsize}{!}{\includegraphics{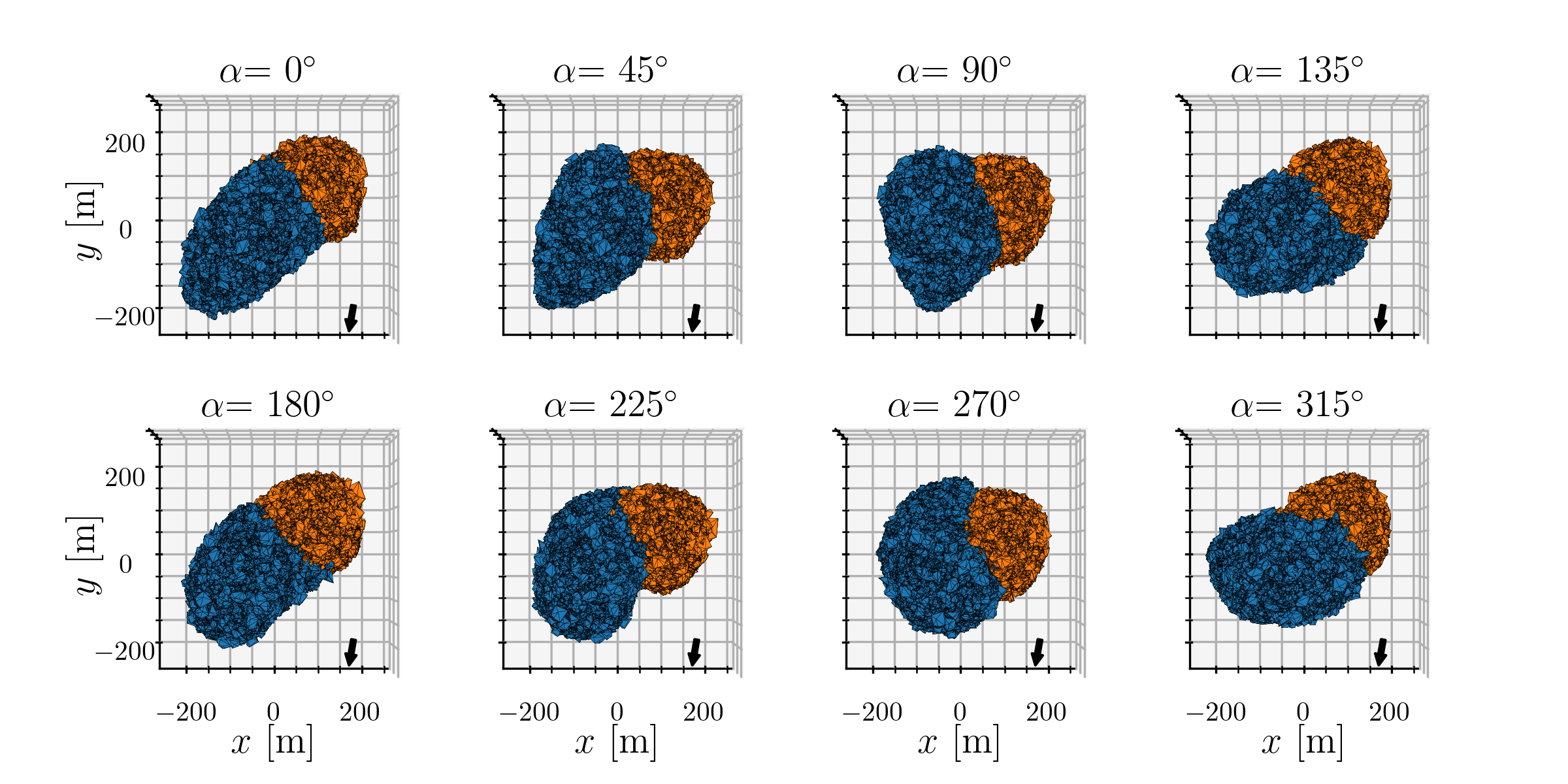}}
    \caption{Higher resolution versions of a few of the cases given in Fig.~\ref{fig:merged_moon_3h}, with the moonlets consisting of 35\,729 particles in total. The moonlet properties are given in Table~\ref{tab:moon_properties_highres}.}
    \label{fig:merged_moon_3h_highres}
\end{figure}

\begin{figure}
    \centering
    \resizebox{0.8\hsize}{!}{\includegraphics{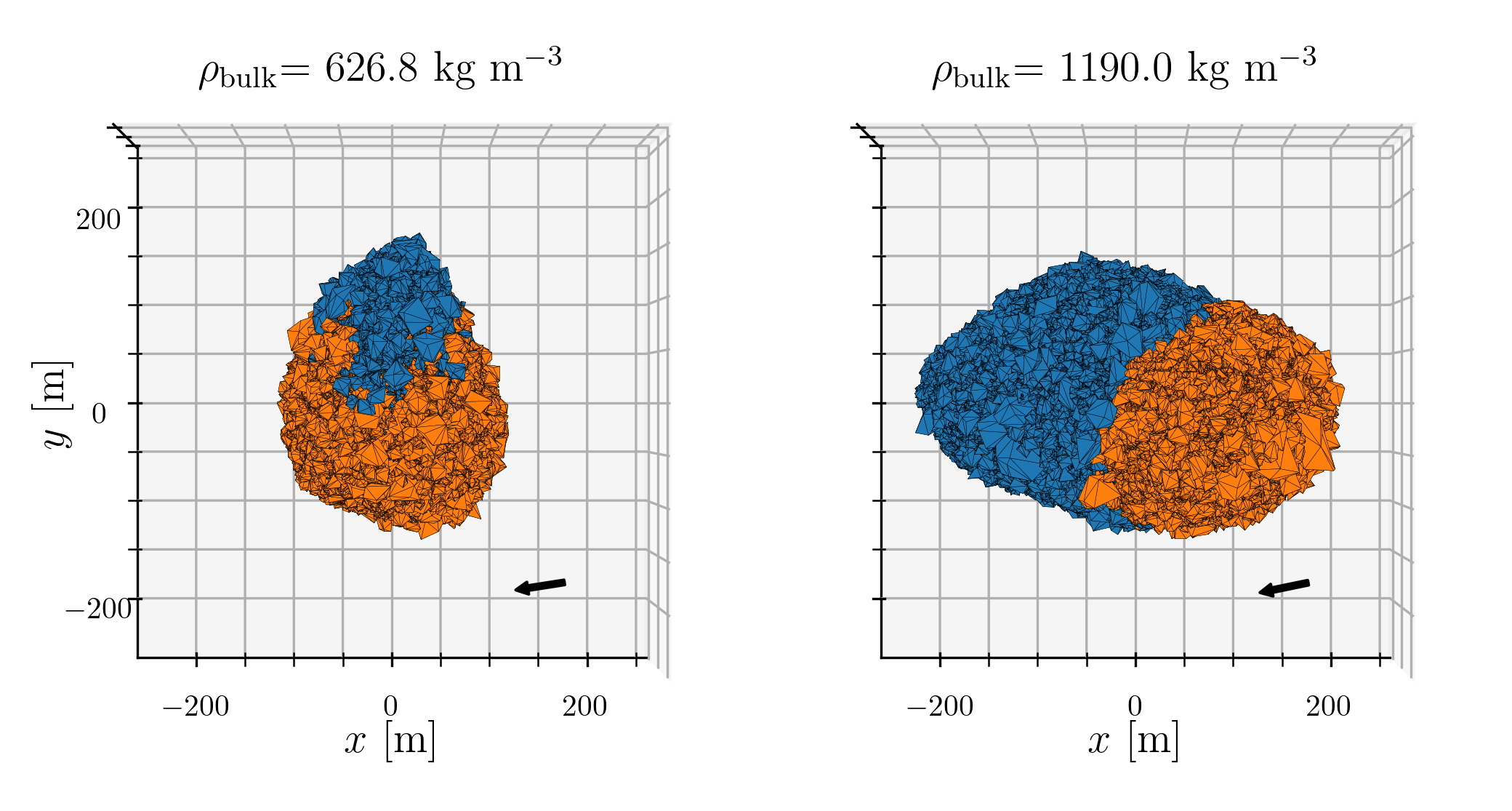}}
    \caption{Final shapes for the case of $\alpha = 0^\circ$ in Fig.~\ref{fig:merged_moon_3h_highres} after an additional 21 h of simulation (25 h in total) for a bulk density of 626.8 kg m$^{-3}$ (left) and 1190 kg m$^{-3}$ (right).}
    \label{fig:merged_moon_25h_highres_alpha0}
\end{figure}

\begin{table}[t]
    \caption{Principal axes and minimum macroscopic porosity data after 3 h of simulation for the lower resolution mergers of Fig.~\ref{fig:merged_moon_3h} and high-resolution cases in Fig.~\ref{fig:merged_moon_3h_highres}. Subscripts of `lr' indicate the corresponding values for lower-resolution moonlets from Table~\ref{tab:moon_properties_48h_1p0} for comparison.}
    \label{tab:moon_properties_highres}
    \centering
    \resizebox{\hsize}{!}{\begin{tabular}{l l l l l l l l l}
        \hline\hline
         $\alpha$ [$^\circ$] & $a_\mathrm{lr}/b_\mathrm{lr}$ & $b_\mathrm{lr}/c_\mathrm{lr}$ & $a_\mathrm{lr}\ [R_\mathrm{primary}]$ & $\phi_\mathrm{lr}$  & $a/b$ & $b/c$ & $a\ [R_\mathrm{primary}]$ & $\phi$ \\
         \hline   
        0 & 2.45 & 0.97 & 0.51 & 0.36 & 2.08 & 1.04 & 0.51 & 0.36\\ 
        45 & 1.82 & 1.25 & 0.46 & 0.34 & 1.61 & 1.21 & 0.45 & 0.35\\ 
        90 & 1.92 & 0.68 & 0.39 & 0.33 & 1.73 & 0.73 & 0.40 & 0.35\\ 
        135 & 1.83 & 1.30 & 0.46 & 0.29 & 1.66 & 1.22 & 0.46 & 0.35\\ 
        180 & 2.50 & 0.83 & 0.49 & 0.33 & 2.13 & 0.91 & 0.49 & 0.34\\ 
        225 & 1.90 & 1.13 & 0.47 & 0.33 & 1.62 & 1.17 & 0.45 & 0.34\\ 
        270 & 1.35 & 1.42 & 0.39 & 0.34 & 1.26 & 1.33 & 0.40 & 0.35\\ 
        315 & 1.67 & 1.28 & 0.44 & 0.29 & 1.61 & 1.18 & 0.45 & 0.36\\ 
         \hline 
    \end{tabular}}
\end{table}

Using the GPU-accelerated Barnes-Hut scheme of \texttt{GRAINS}, we carried out eight simulations for the moonlets in Table~\ref{tab:moonlet_initial_data_highres} with $\alpha$ increments of 45$^\circ$ up to the 3 h mark, as these simulations are computationally heavy with $N_\mathrm{p}=35\,729$, excluding the primary (for visualisations of a few different mergers, see supplementary movies\footnote{The movies are titled \texttt{alphaXdeg\_merger\_highres.avi}, where X represents an angle $\alpha\in\{0,45,90,135,180,225\}^\circ$.}). The initial bulk density of the cut-out aggregates did not match the target density of 626.8 kg m$^{-3}$. Therefore, we had to scale the aggregate properties, which yielded a corresponding material density of 1134 kg m$^{-3}$. The aggregates formed from each merger are shown in Fig.~\ref{fig:merged_moon_3h_highres} and Table~\ref{tab:moon_properties_highres} shows the principal axes data, as well as porosities, for both the low- and high-resolution scenarios. Both from a visual comparison with Fig.~\ref{fig:merged_moon_3h} and the principal axes values, we observe a significant change in aggregate structure. The steeper SFD indeed serves to smooth out contacts between larger boulders, generally leading to less elongated shapes and more homogeneous surfaces. With an increase in resolution by a factor of almost 20, this also greatly contributed to the homogeneity of the surface. Furthermore, the estimate for our DEEVE values also becomes more accurate with larger values of $N_\mathrm{p}$. Together with the smoother, less elongated shapes, this helps explain the discrepancy in $a/b$ ratios for the objects, while the principal axes $a$ remain largely similar. Yet, despite the reduced rigidity of the aggregate, the sub-escape-velocity nature of the mergers still leads to clear bilobate shapes. We can distinguish necks for the LAB cases $\alpha = 135$ and 180$^\circ$ and the OAB shape $\alpha = 315^\circ$, as well as SAB structures for $\alpha= 45,\ 90,\ 225$ and $270^\circ$. Moreover, after dynamically propagating the $\alpha = 0^\circ$ system in time for an additional 22 h at $r_0$ (25 h in total), we saw a similar catastrophic tidal disruption to the low-resolution case (shown in movie \texttt{alpha0deg\_post\_merger\_highres\_627kgm3.avi}), leading to a largest remnant on an eccentric orbit with $m_\mathrm{moon}=0.40M_0$, $a/b = 1.30$, $b/c = 1.15$ and $a = 0.27R_\mathrm{primary}$, as seen in the left image of Fig.~\ref{fig:merged_moon_25h_highres_alpha0}. We also ran a comparison test for the same orbital properties and shape but higher bulk and particle densities of $\rho_\mathrm{bulk}=1190$ kg m$^{-3}$ and $\rho_\mathrm{p}=2154$ kg m$^{-3}$, giving the moon the same bulk density as the primary, which is predicted to be the case for the Didymos system \citep{Daly_et_al_2023a}. In this scenario, the moon did not undergo any tidal disruption event and retained all its mass (see movie titled \texttt{
alpha0deg\_post\_merger\_highres\_1190kgm3.avi}). Deformation primarily occurred in the form of additional displacement of boulders. The final, more homogeneous shape can be seen in the right plot of Fig.~\ref{fig:merged_moon_25h_highres_alpha0} and has the properties: $a/b=1.60$, $b/c=1.05$, $a = 0.41R_\mathrm{primary}$ and $\phi = 0.29$. Tidal homogenisation and straightening made the body more elongated, similar to a prolate spheroid with an even less distinguishable boundary between the two moonlets, only evident from a shallow neck that can be detected when studying its $\alpha$-wrap model.

\section{Discussion}\label{section:discussion}

\begin{figure}
    \resizebox{0.8\hsize}{!}{\includegraphics{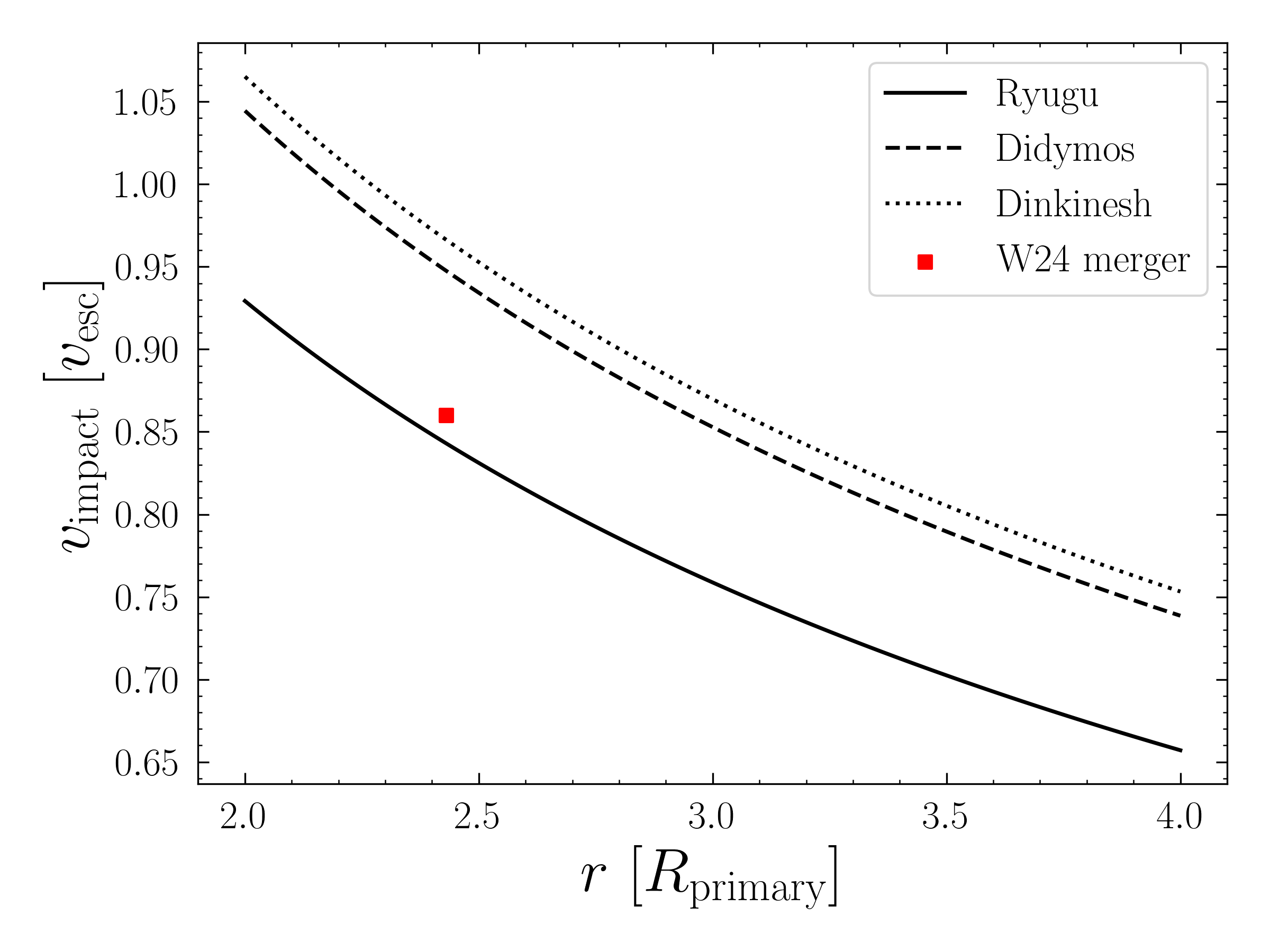}}
    \caption{Impact velocity in terms of mutual escape velocities for the case of the merger data in Table~\ref{tab:W24_merger_data} and moonlets in Table~\ref{tab:moonlet_initial_data}, setting a static bulk density of 1190 kg m$^{-3}$. The red square represents the values for the \citetalias{Wimarsson_et_al_2024} merger.}
    \label{fig:impact_velocities_vs_dist}
\end{figure}

The variety in shapes obtained from small adjustments of our initial conditions reinforces our understanding that forming binary asteroid systems through the debris disk model is a highly chaotic and unpredictable process \citepalias{Agrusa_et_al_2024,Wimarsson_et_al_2024}. Notably, most of the pre-impact properties of the W24 merger, such as the mass ratio, impact velocity, distance and angle, along with the moonlet shapes, were kept intact, with only the orientation of the most massive moonlet and particle SFD being altered. In turn, we have only explored a small portion of the extensive parameter space for mergers in the sub-escape-velocity regime. Looking at the distribution of properties for the major mergers observed in the \citetalias{Wimarsson_et_al_2024} simulations, the most massive moonlet is always more or less prolate in nature and dominant in mass compared to the second (see their Sect.~5.2), with the chosen merger case for this study being close to the median values in both elongation and mass ratio. As for impact velocity, it is one of the most extreme cases with a value 0.84$v_\mathrm{esc}$, since it occurs closer to the primary than a majority of the other mergers. Despite this fact, we still see clear bilobate structures, even for the high-resolution simulations. Hence, a more typical impact velocity, near the median value of ${\sim}0.7 v_\mathrm{esc}$ (based on the sample size of major mergers in \citetalias{Wimarsson_et_al_2024}) would result in even less deformation and more distinguishable bilobate features. In Fig.~\ref{fig:impact_velocities_vs_dist}, we have plotted the scaling of the impact velocity with distance from the primary for the W24 merger case in Table~\ref{tab:W24_merger_data}. For the case of simplicity, we have assumed the largest moonlet to be on a circular orbit and scaled the velocity of the second moonlet according to their relative magnitudes and impact angle, i.e.~rotating the velocity vector of the second, impacting moonlet in the direction of the primary relative to the circular orbit velocity. Given the low bulk densities of our moonlets, here set to a static 1190 kg m$^{-3}$, we assume that the impact velocity is comparable to the relative velocity \citep{Leleu_et_al_2018}. The three different graphs each represent a rubble-pile primary, all assumed to be spherical in nature, with the fiducial case of Ryugu in solid having $\rho_\mathrm{mean}=1190$ kg m$^{-3}$ and $R_\mathrm{primary} = 500$ m \citep{Barnouin_et_al_2019}. The dashed line is representing Didymos with $\rho_\mathrm{mean}=2400$ kg m$^{-3}$ and $R_\mathrm{primary} = 382.5$ m \citep{Daly_et_al_2023a}, while the dotted line shows the velocities for Dinkinesh with $\rho_\mathrm{mean}=2700$ kg m$^{-3}$ and $R_\mathrm{primary} = 395$ m \citep{Levison_et_al_2024}. It becomes clear that even for this extreme case with an impact angle of 21.5$^\circ$ occurring so close to the primary, we still get the sub-escape-velocity mergers for a much denser main body.

In turn, while we lack sufficient information regarding the full catalogue of potential shapes formed through sub-escape-velocity mergers, the samples from Figs.~\ref{fig:merged_moon_3h}, \ref{fig:merged_moon_48h_1p1} and \ref{fig:merged_moon_3h_highres} cover a significant fraction of them and identifies that observed contact binary shape types such as LAB (Selam-like), OAB (Itokawa-like) and SAB (Apophis-like) could be formed in orbit around rubble-pile primaries similar in size and density to Ryugu, Bennu, Didymos and Dinkinesh.

\subsection{Post-merger evolution and tidal distortion}\label{section:discussion_tidal_study}

\begin{figure}
    \resizebox{\hsize}{!}{\includegraphics{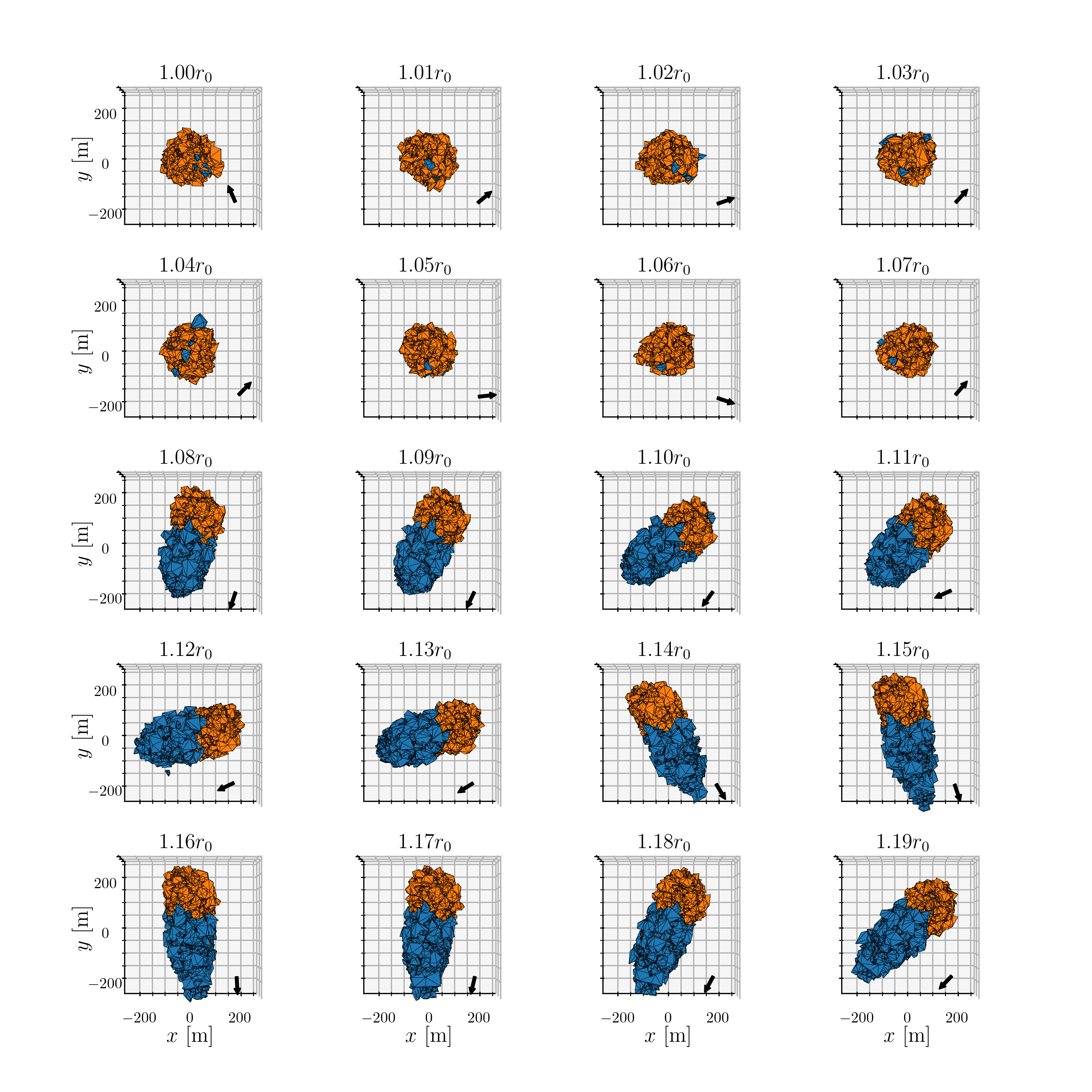}}
    \caption{Shapes and orientations of aggregates after 45 h of post-merger evolution, based on the case of $\alpha = 0^\circ$ in Fig.~\ref{fig:merged_moon_3h}. Each case represents a different shift in distance from the primary barycenter after the merger has occurred, between $r_0$ and $1.19r_0$.}
    \label{fig:merged_moon_48h_norm_dist}
\end{figure}

\begin{figure}
    \centering
    \resizebox{\hsize}{!}{\includegraphics{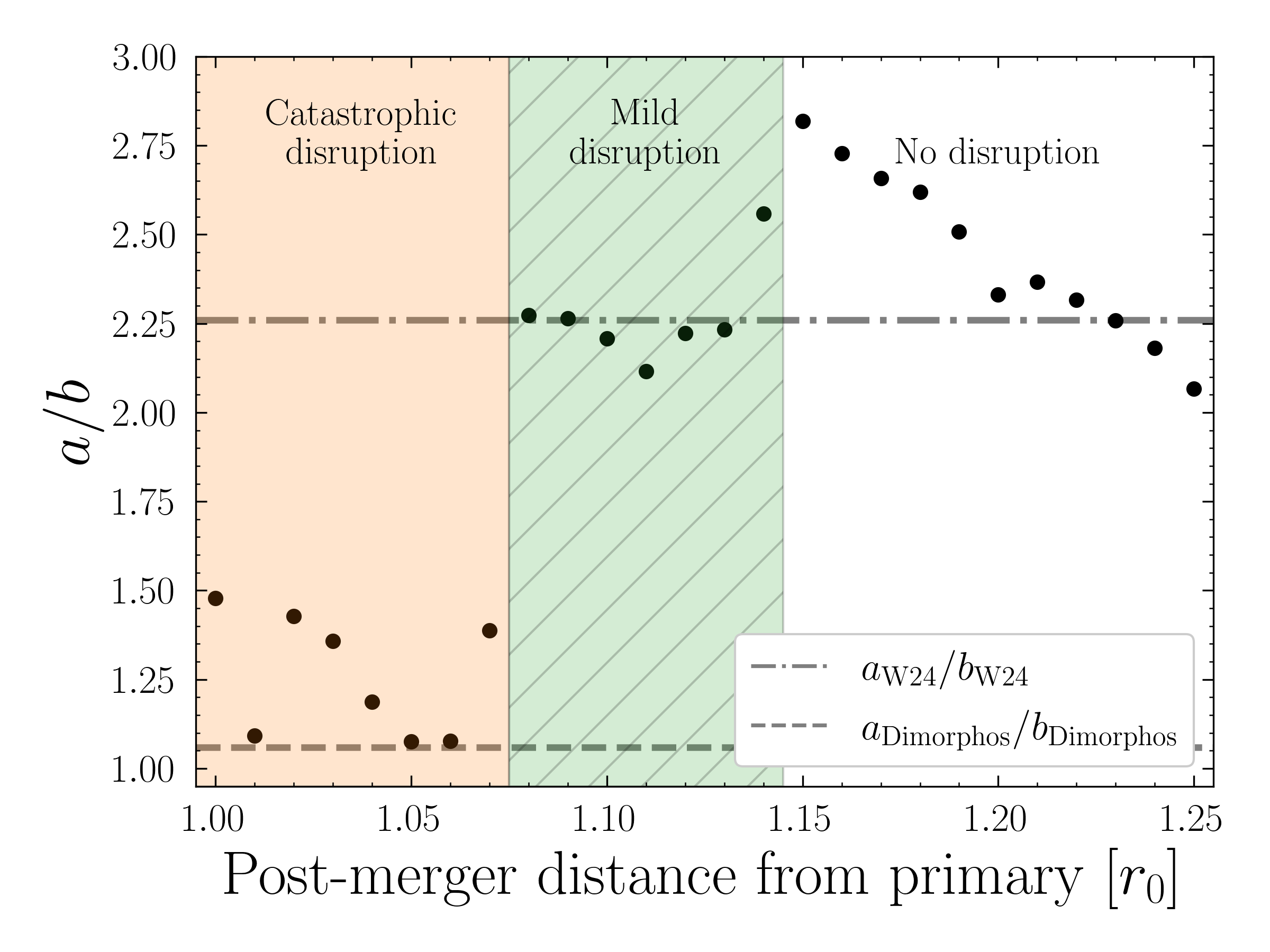}}
    \caption{Principal axis ratio $a/b$ for the largest remnant after 45 h of simulation at different shifted distances from the primary barycenter post merger (in $r_0=2.43R_\mathrm{primary}$) for the case of $\alpha = 0^\circ$. The dashed line shows the estimated $a/b$ value of Dimorphos \citep{Daly_et_al_2024} while the dashed-dotted line shows the final $a/b$ value for the reference W24 merger, evolved with the debris disk still present.}
    \label{fig:ab_vals_norm_dist}
\end{figure}

\begin{figure}
    \centering
    \resizebox{\hsize}{!}{\includegraphics{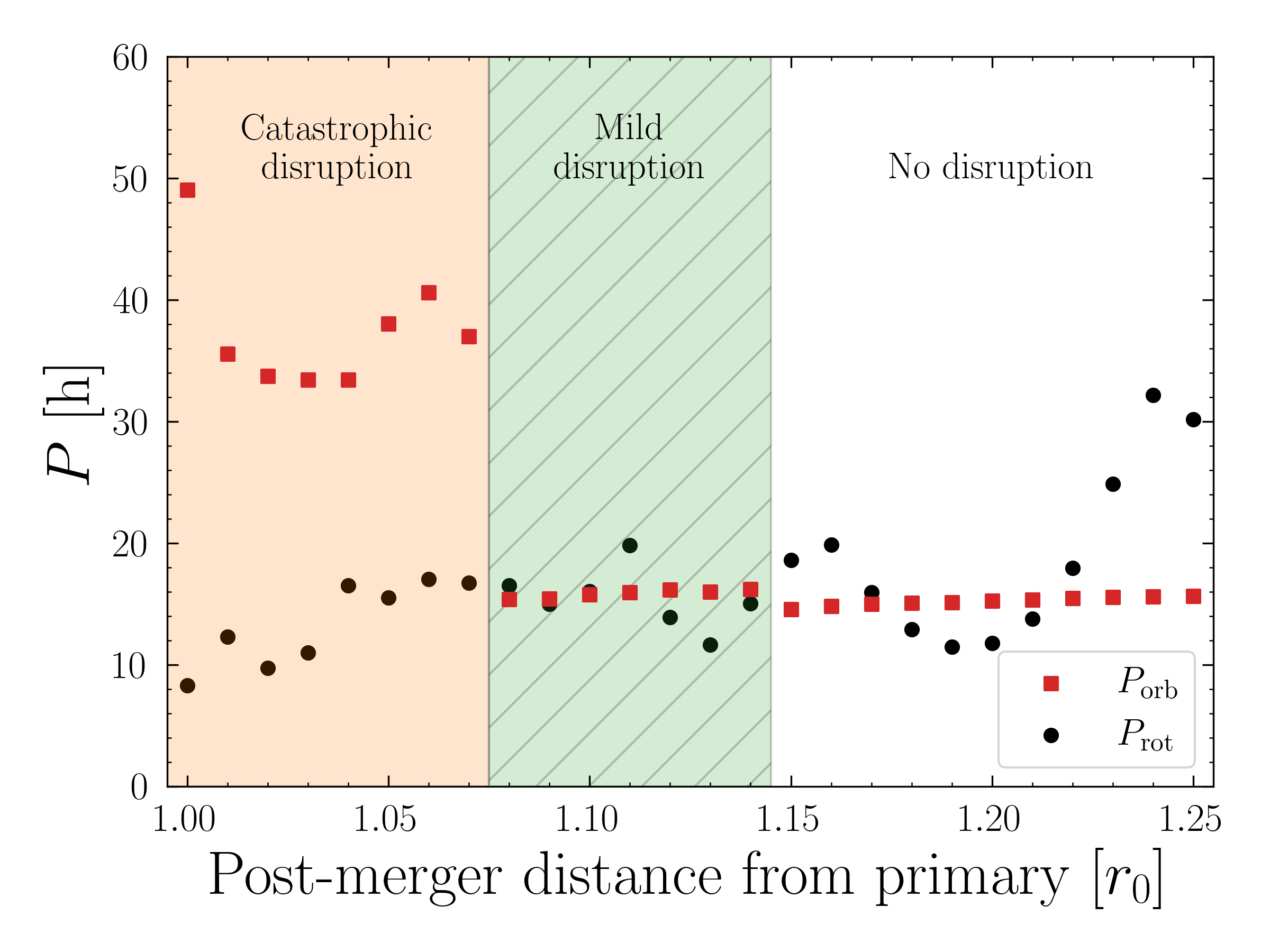}}
    \caption{Orbital (red square) and rotational (black dot) periods for the largest remnant after 45 h of simulation (48 h total) when initiated at different shifted distances from the primary barycenter for the case of $\alpha = 0^\circ$.}
    \label{fig:periods_norm_dist}
\end{figure}

After formation, the moons show susceptibility to deformation due to tidal forces from the primary via disruption and mass-shedding or displacement of boulders, leading to homogenisation, elongation or all of them. Having explored three cases for the configuration in Tables~\ref{tab:W24_merger_data} and \ref{tab:moonlet_initial_data} with varying distance from the primary barycenter, $r_0 = 2.43R_\mathrm{primary}$, $1.1r_0 = 2.673R_\mathrm{primary}$ and $1.2r_0 = 2.916R_\mathrm{primary}$, we observe variations in the level and nature of deformation, also for different shape types. Where several of the more initially elongated bilobate satellites fail structurally for the $1.1r_0$ case in Fig.~\ref{fig:merged_moon_48h_1p1} and undergo catastrophic disruption, other shapes retain all their mass. Hence, using analytical predictors for the tidal behaviours of these objects does not suffice, as initial shape and state of rotation alter the outcome even at similar distances from the primary, which is in agreement with the findings of \citet{Burnett_Fodde_&_Ferrari_2025}, who investigated tidal dissipation in rubble-pile asteroid satellites. To better our understanding of the tidal disruption regimes for a given shape and particle composition, we performed a set of 45 h simulations for the case of the nominal $\alpha = 0^\circ$, where we altered its distance from the primary in increments of 0.01$r_0$ between $1.01r_0$ and $1.25r_0$. The resulting aggregates at the final time step for cases initiated up to $1.19r_0$ from the primary have been plotted in Fig.~\ref{fig:merged_moon_48h_norm_dist}. The geometrical and orbital properties of the largest remnant at the end of each simulation are provided in Table~\ref{tab:moon_properties_48h_norm_dist}. 

\begin{figure}
    \centering
    \resizebox{\hsize}{!}{\includegraphics{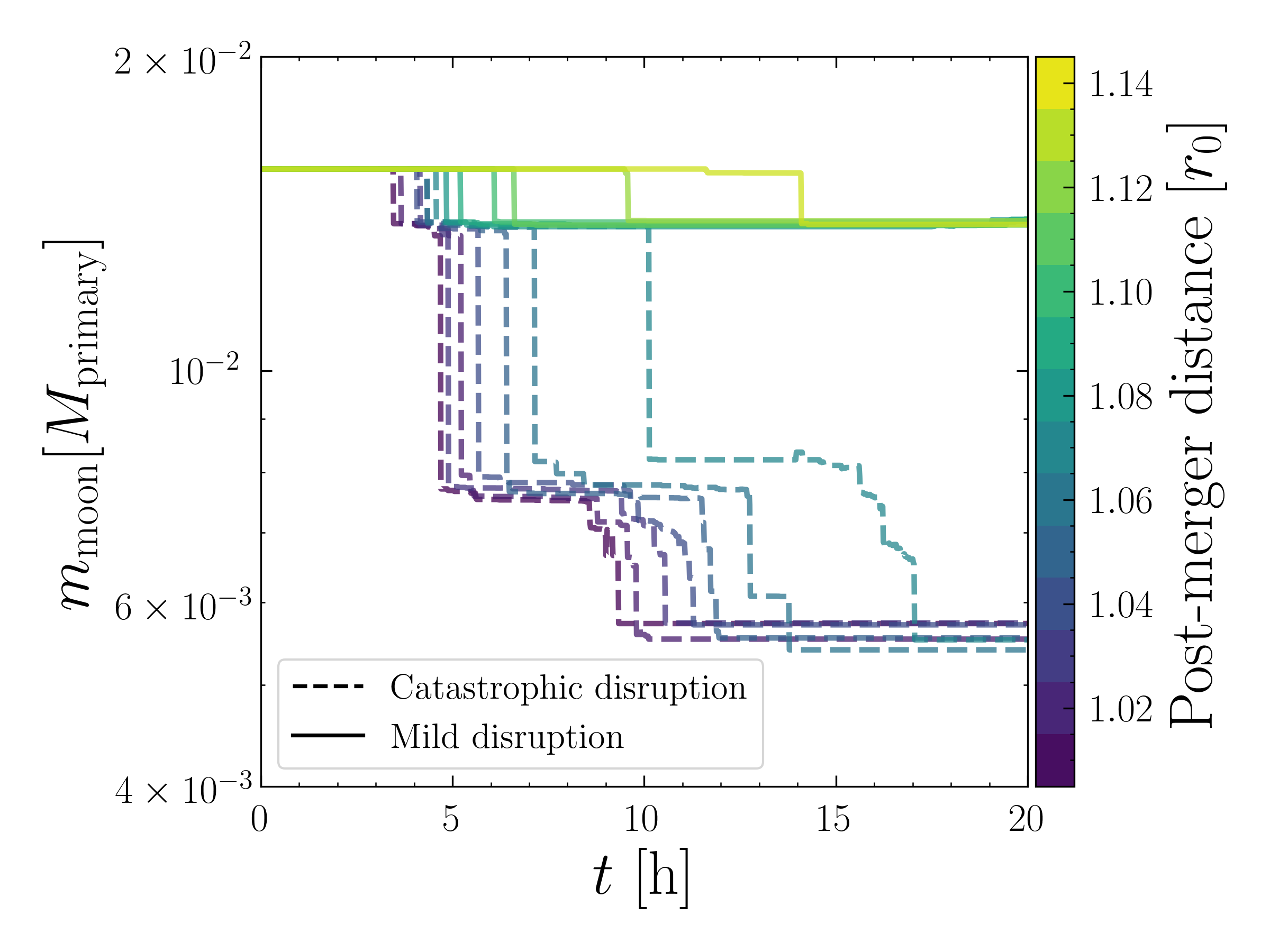}}
    \caption{Change in mass over time for the simulation scenarios leading to the shapes in Fig.~\ref{fig:merged_moon_48h_norm_dist} that undergo tidal disruption. The bodies undergoing catastrophic disruption have dashed lines, while the mild disruption scenarios have solid lines. The colour of each line represents its initial post-merger distance from the primary.}
    \label{fig:mass_loss_norm_dist}
\end{figure}

\begin{figure}
    \centering
    \resizebox{\hsize}{!}{\includegraphics{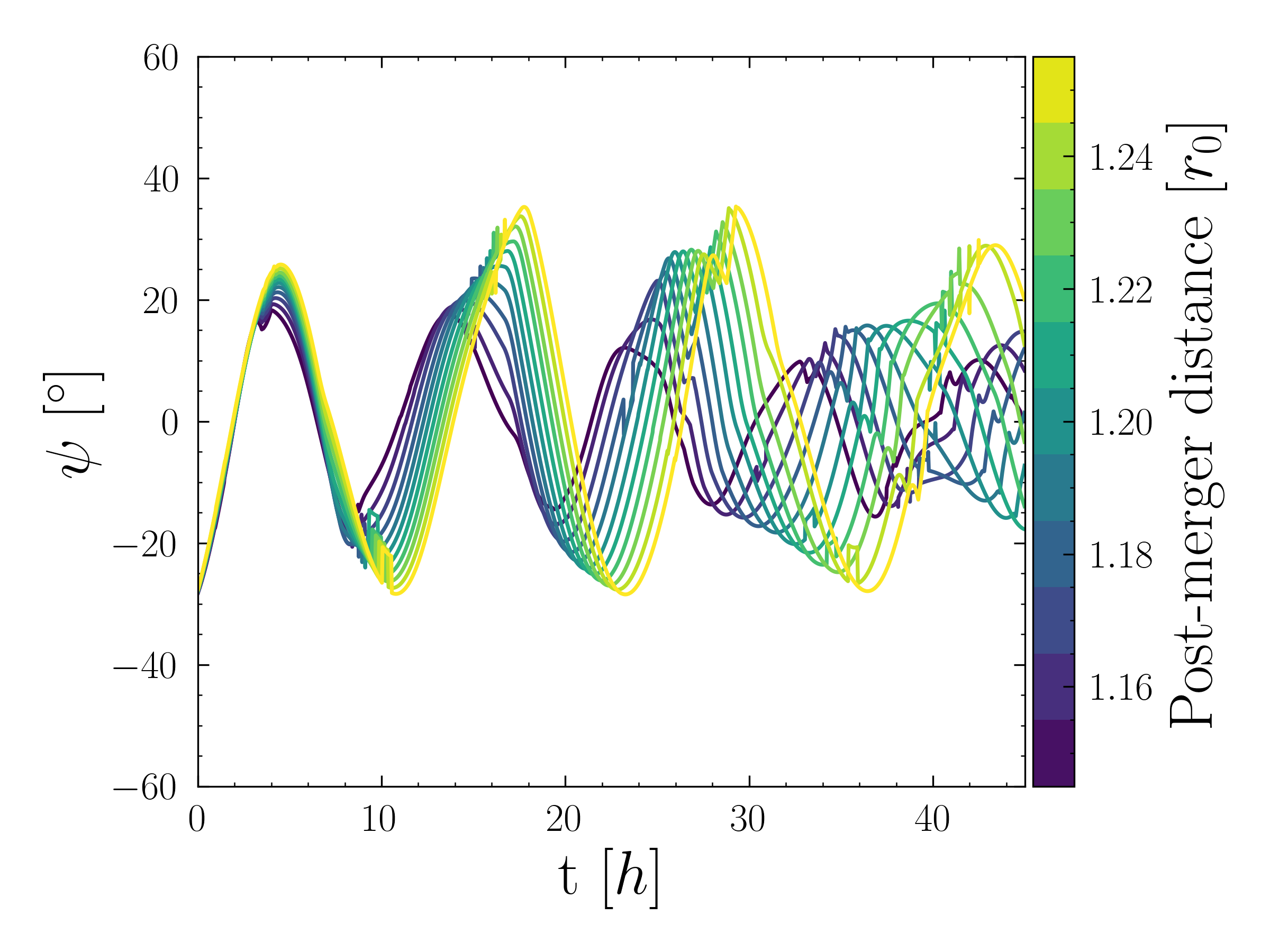}}
    \caption{Post-merger change in yaw angle over time for the no-disruption cases of Fig.~\ref{fig:ab_vals_norm_dist}. The colour of each line represents the initial distance in $r_0=2.43R_\mathrm{primary}$. The data has been smoothed using a cumulative moving average with a window width of 30 time steps to reduce noise caused by small shifts in the body-fixed frame.}
    \label{fig:yaw_angle_at_mass_loss_vs_time_norm_dist}
\end{figure}

Notably, there seems to exist a discrete distance where the initially bilobate structure of the aggregate can stay intact. Inside this distance, nearly all particles originally belonging to moonlet A get tidally stripped from the moon, leaving the original moonlet B behind with a few additional grains that are either retained or re-accreted. From the distance of $1.08r_0$ to $1.14r_0$, only about 10\% of moonlet A is shed, while the formed moon retains all of its mass from $1.15r_0$ and onward. This phenomenon can be seen in Fig.~\ref{fig:ab_vals_norm_dist}, which shows the $a/b$ values for each moon at the end of the simulations. At post-merger distances of $r<1.08r_0$, the moon undergoes a catastrophic tidal disruption event, leaving behind a Dimorphos-like OS on an eccentric orbit, $e_\mathrm{orb}\gtrsim 0.4$. Once more, we see examples of how our DEEVE estimates for $a/b$ fail for very small bodies, as all these objects exhibit clear OS shapes from a visual evaluation, but only three of them have values matching that of Dimorphos. When instead only 10\% of the mass is lost in a mild tidal disruption event, we see a sharp increase in $a/b$, as the final body highly resembles the reference W24 merger, which evolved at $r_0$ with the disk still present. Even for the same type of disruption, there are distinct variations in their geometric features, as the moons evolved at $1.08r_0$ and $1.09r_0$ exhibit LAB features. Further out from the primary, the objects become more compact and homogeneous with a dip in $a/b$ for $1.11r_0$, before the bilobate structure re-emerges at $1.13r_0$. To understand this trend, we take a look at the rotational and orbital periods at 48 h for all moons, shown in Fig.~\ref{fig:periods_norm_dist}. Here, the $y$-axis shows the periods at the end of each simulation for different initial post-merger distances from the primary. Note that the orbital periods are highly similar for cases that undergo mild and no disruption because the former migrated outwards after losing mass (see Table~\ref{tab:moon_properties_48h_norm_dist}). With the reduction in $a/b$ in Fig.~\ref{fig:ab_vals_norm_dist}, there is a corresponding increase in rotational period in Fig.~\ref{fig:periods_norm_dist}, indicating that a non-synchronous rotational state leads to a less prolate shape. From Table~\ref{tab:moon_properties_48h_norm_dist}, we observe that it is mainly changes in $b$ that cause this decrease, as both $a$ and $\phi$ stay similar compared to other cases in this regime. The changes in $b$ can be attributed both to tidal distortion in the form of stretching and re-accretion of shed material along this principal axis. Analysing the change in mass over time for each tidal disruption scenario in Fig.~\ref{fig:mass_loss_norm_dist}, we also see how the moons undergo mass-loss at different points in their dynamical evolution, showing the distance-dependence of the tidal strength. Given that all bodies have the same initial angular velocity, this comes from the fact that each moon is at a different stage of its rotation when the mass-loss occurs. This, in turn, is highly dependent on the level of tidal dissipation in the aggregate material, which greatly decreases with distance from the primary. Studying the yaw angle, $\psi$, (computed as the rotation around the principal axis in the equatorial plane of the primary) at the time step where each body first undergoes mass-loss, we see a clear pattern where the bodies at $1.08r_0$ and $1.09r_0$ have $\psi$-values closer to zero than for the subsequent cases (see Fig.~\ref{fig:yaw_angle_at_mass_loss_norm_dist} in appendix \ref{appendix:plots}). Hence, the tidal forces are only strong enough at distances $1.06r_0\leq r \leq 1.10r_0$ to synchronise the rotation of the objects on the timescales considered here (48 h), causing the pull to be parallel to the longest principal axis of the aggregate, $a$, leaving it elongated and creating a distinct neck between the lobes. In Fig.~\ref{fig:yaw_angle_at_mass_loss_vs_time_norm_dist}, we show the change in yaw over time for no-disruption scenarios. An initial offset of the libration amplitude is expected, due to the rotational period being kept static while the orbital period changes with distance, but the evolution over time is non-linear, highlighting the importance of tidal dissipation for the orbital and structural evolution of each moon.

At $1.14r_0$, the elongation of the moon becomes more pronounced again as tides are strong enough to stretch out the body, while also stripping off some mass. The $a/b$ value reaches its peak at $1.15r_0$, where all mass is retained, but tidal forces are still sufficiently strong to elongate the body. Beyond this point, tides gradually grow weaker but also act at larger angles relative to the principal axis, $a$. The change in libration amplitude with distance over time reinforces the complicated nature of the aforementioned spin-state of each moon and how it depends on tidal disruption history and tidal strengths at the new post-merger positions of the moons. Although the exhibited pattern calls for deeper investigation beyond the scope of this work, the sensitivity of the spin-orbit state points towards the high tidally dissipative nature of rubble-pile satellites indicated from the results of \citet{Burnett_Fodde_&_Ferrari_2025}. Crucially, from this analysis, we learn that the sub-escape-velocity nature of mergers in this regime highly affects the final moon shapes. Due to the soft impact and rigidity of our moonlets, there is limited initial displacement of boulders. Hence, structural failure always occurs at the same places, the innermost edge of moonlet A closest to the primary and the connection point between the two moonlets. This leads to a predictable behaviour, where the entirety of moonlet A is stripped off at $r<1.08r_0$ and only the tip of moonlet A is lost at $1.08r_0\leq r < 1.15r_0$. In the latter region, the tides also serve to create the aforementioned distinct neck between the two moonlets, most visible in the cases $1.08r_0$, $1.09r_0$ and $1.13r_0$. We further note that distinct necks can also form without the moon undergoing any tidal disruption events, as the case of $\alpha=225^\circ$ in Figs.~\ref{fig:merged_moon_3h} and $\ref{fig:merged_moon_48h_1p1}$ where an OAB structure has been reshaped into a LAB without any mass-loss involved.

\subsection{Particle number and size distribution}

With the considerable improvement in resolution achieved with the GPU-acceleration scheme of \texttt{GRAINS}, increasing the particle number by almost a factor of 20, we could more closely study the effect of having a steeper SFD, leading to a considerably larger fraction of smaller particles compared to the relatively low-resolution configuration of \citetalias{Wimarsson_et_al_2024}. At a first glance, coarser, lower resolution distributions offer an approximation of deformation, as we see clear similarities when comparing the shapes in Figs.~\ref{fig:merged_moon_3h} and \ref{fig:merged_moon_3h_highres}. Moreover, it suffers comparable levels of disruption to our high-resolution moons, where most of moonlets A and A2 are lost in the respective cases. However, the increase in stability where the largest remnant of the high-resolution aggregate retains more mass shows the importance of resolving particle--particle contacts on a smaller scale when studying tidal deformation. 

Recent studies on the behaviour of granular media in tidal disruption scenarios have primarily investigated the effects of altering particle distribution properties for rubble piles consisting of spherical elements \citep{Zhang_&_Lin_2020,Zhang_&_Michel_2020}, approximating granular contact mechanisms with a linear spring-dashpot system along with a non-spherical shape parameter \citep{Schwartz_et_al_2012,Zhang_et_al_2017}. The few studies that have used polyhedral particles have either opted for a monodisperse \citep{Marohnic_et_al_2023} or uniform SFD \citep{Movshovitz_et_al_2012}. It is difficult to meaningfully compare our results with the aforementioned DEM studies because of the spread in: problem scale, ranging from encounters with stars and terrestrial planets to small asteroids, the variation in particle size distributions, shapes and packing, as well as the contact methods and the numerical capabilities available at the time each study was conducted. Even so, we will address a few points and key differences that could prove useful for future investigations of rubble-pile stability. Firstly, while brought up as likely to be important, studying the effect of particle SFD for rigidity was beyond the scope of these works. Since then, a series of articles using SPH methods have found that boulder packing and size distribution can have a crucial effect on the stability of rubble-pile objects during high-energy collisions \citep{Raducan_et_al_2022,Raducan_et_al_2024a,Raducan_et_al_2024b}. Moreover, for our purposes, observations of Dimorphos's surface indicate a boulder SFD with a mean and maximum diameter of 1.4 and 16 m, respectively, with a best fit power-law index of $-3.4\pm 1.3$ aligning with similar observed SFDs from Didymos, Itokawa, Eros and Toutatis, which all have indices below $-3$ \citep{Pajola_et_al_2024}. Secondly, we once more emphasise that rubble-pile satellites in binary asteroid systems are likely more dissipative than previously believed \citep{Burnett_Fodde_&_Ferrari_2025} and that polydisperse SFDs can facilitate propagation of mechanical waves \citep{Marti_et_al_2024} in granular media and reduce macroporosity due to improved packing \citep{Ferrari_&_Tanga_2020}. These conclusions align with our results, as there is an evident change in the amount of deformation that occurs during mergers for our high-resolution case. Thirdly, we highlight the fact that the mentioned studies have been focusing on disruptions caused by close encounters with stars or planets. It might be that the difference in scale explains why they detected a seemingly weaker dependence on resolution for aggregate stability. The consequent improvement in stability caused by increasing $N_\mathrm{p}$ from $10^3$ to $10^4$ is possibly too weak relative to the strength of the tidal forces in these environments, while highly relevant when the primary is a small asteroid. A difference in mass-loss of $0.05M_0$ when comparing the low- and high-resolution disruptions is small in the scope of the system. Yet, this amount of mass can be decisive when attempting to accurately capture deformation. Finally, it is critical to acknowledge the uniqueness of the structure considered in our tidal study. Not only is it highly elongated, but it is also of a bilobate nature with a weak neck. Investigating a more structurally homogeneous target, we would likely not have identified the notable shift in dynamical behaviour under the influence of tidal forces.

Thus, we argue that it is crucial that DEM studies use models with polydisperse particle distributions and methods that can capture the mechanical effects of irregularly shaped grains to get a full picture of the reshaping of rubble-pile targets in asteroid environments from mergers and tidal forces. With all the complexities at hand, this calls for a deeper investigation beyond the scope of this work into the effects of varying particle SFDs, aggregate bulk density and packing.

\subsection{Sub-escape-velocity merger formation scenarios for Dimorphos and Selam}

While we still have limited direct observations of asteroid satellites, the recent imaging of Dimorphos by the DART spacecraft \citep{Daly_et_al_2023a,Chabot_et_al_2024} and Selam by Lucy \citep{Levison_et_al_2024} has provided us with invaluable information. Given that they both orbit small, rapidly rotating top-shaped asteroids, their existence agrees with a debris disk formation scenario. From the results and analysis in \citetalias{Wimarsson_et_al_2024}, there were strong hints that sub-escape-velocity mergers can generate both oblate and bilobate shapes. This has been further reinforced in this work. Additionally, we have shown that OS and Selam-like LAB satellites do not necessarily need to form directly from mergers, as they can end up with similar shapes from post-merger dynamical events. Dimorphos-like objects can consistently form from catastrophic tidal disruptions, as long as the merger occurs sufficiently close to the primary. In our simulations, these objects end up on eccentric orbits. However, given the influence of a disk on the orbital evolution of our moons (see Fig.~\ref{fig:disk_vs_no_disk}), we cannot exclude the possibility that the presence of a disk can circularise the orbit. Moreover, from Fig.~\ref{fig:periods_norm_dist} we see that the final rotation of the largest remnant in these catastrophic disruption events is dependent on whether it occurred when its progenitor was tidally locked with the primary. From measurements of Dimorphos's rotational state before the DART impact, it appeared to be in a synchronous state \citep{Rivkin_et_al_2021,Agrusa_et_al_2022,Naidu_et_al_2024}. For the case of e.g.~$1.05r_0$, the rotational period of the body appears to match the corresponding period for a circular orbit at ${\sim}3R_\mathrm{primary}$, which is where Dimorphos is situated in the Didymos system. Hence, disregarding any secular synchronisation of the rotation for a moment, given that the tidal disruption distance would be dependent on the progenitor's bulk density, we should be able to predict where a possible merger and subsequent disruption must have occurred for the satellite to have formed via this scenario. This would then be made possible with improved measurements of Dimorphos's internal properties from the Hera mission \citep{Michel_et_al_2022}. Numerical experiments of impacts into a Dimorphos-like target have further found that the bulk density of the object is likely lower than that of Didymos \citep{Raducan_et_al_2024b}, which also speaks for this tidal disruption formation model, as this would shift the disruption limit inwards, closer to the primary.

Predicting the formation history of Selam is a more challenging task. The contact binary satellite is surprisingly massive with an estimated $m_\mathrm{moon}=0.06M_\mathrm{primary}$ and orbits Dinkinesh with a semi-major axis of $9R_\mathrm{primary}$. What is more, the geological features are highly distinct, with a narrow neck connecting two equal-sized lobes with diameters of 210 m and 230 m \citep{Levison_et_al_2024}. One of the lobes has a noticeable ridge that appears to span around it, inclined at an angle of ${\sim}50^\circ$ relative to the orbital plane. It was suggested by \citet{Levison_et_al_2024} that this ridge could have been formed by accretion of material from a debris disk. Alternatively, it could have formed by mergers of similar-sized moonlets \citep{Raducan_et_al_2025}. In order to produce Selam from a single mass-shedding event, the debris disk must be of substantially higher mass compared to the cases explored in \citetalias{Wimarsson_et_al_2024} and \citet{Agrusa_et_al_2024}. Its contact binary structure could then be formed via a mild tidal disruption after most of the disk mass has been bound in a single, merged moon similar to our $\alpha=0^\circ$ case. From our analysis in Sect.~\ref{section:discussion_tidal_study}, we know that such a bilobate structure would fail at two specific points, the closest point to the primary and the contact area between the moonlets. Hence, there could be a discrete distance, such as $1.08r_0$ in Fig.~\ref{fig:merged_moon_48h_norm_dist} for the $\alpha=0^\circ$ body, at which the tidal forces will strip off the innermost edge of the most massive moonlet, leaving it less prolate while at the same time pulling the lobes apart and creating the neck between them. Such an event is likely to be followed by migration further out into the system, where the satellite would be less susceptible to homogenisation by tides, retaining its LAB shape. It remains to be investigated whether a disruption followed by circularisation via angular momentum exchange with a surrounding disk could be a key mechanism in placing Selam at its current position at $9R_\mathrm{primary}$, or if higher order effects such as binary YORP are necessary for it to migrate this far out \citep{Jacobson_et_al_2014}, which would occur over a period of ${\sim}10^6{-}10^7$ years \citep{Merrill_et_al_2024}. Still, within this single mass-shedding formation scenario, it would be non-trivial to explain the characteristic lobe ridge. We cannot resolve such a feature in our current debris disk simulations, as that would require more particles. To reach impact velocities that could create it via a merger, the impact angle would have to be much larger than those observed in \citetalias{Wimarsson_et_al_2024}. For this to occur, the largest moonlet would have to undergo a tidal disruption, stripping off most of its mass and leaving it less prolate while putting it on a highly eccentric or unbound orbit where it collides with a lower-mass moonlet (which tend to have more oblate structures). It could alternatively be accelerated by some other dynamical effect, such as Lindblad resonance with the remaining disk, as proposed by \citet{Raducan_et_al_2025}. Given that the disks in \citetalias{Wimarsson_et_al_2024} and \citet{Agrusa_et_al_2024} fully disperse over just a few days, we argue that this would be an unlikely mechanism in our model. For a formation history involving two or more mass-shedding events, we can imagine that two separate OS satellites were generated via tidal disruptions, one-by-one, leading to migration and a ${\sim}v_\mathrm{esc}$ merger beyond $3R_\mathrm{primary}$, similar to the pyramidal regime scenario discussed in both \citet{Levison_et_al_2024} and \citet{Raducan_et_al_2025}. Whichever scenario is true for the mysterious contact binary satellite, it is evident that its dynamical history is highly complex and bound to teach us more about the chaotic nature of binary asteroid system formation.

\section{Conclusions}\label{section:conclusions}

In a first approach to explore which dynamical mechanisms determine moon shapes resulting from sub-escape-velocity mergers, we have performed several sets of polyhedral $N$-body simulations. We find that each moonlet undergoes little initial deformation due to the impact because of its sub-escape-velocity nature. The orientation of the prolate, most massive moonlet in our configuration highly affects the final shape of the formed moon, creating a wide range of post-merger structures which we classify as long-axis bilobate (LAB), off-axis bilobate (OAB) and short-axis bilobate (SAB). Further, we studied the stability of the formed moons, letting them dynamically evolve around their Ryugu-like primary for up to 48 h at various distances from the primary. Depending on their shape and rotational state, the bodies undergo different types of deformation via tidal disruption, as well as tidal distortion, which can significantly alter their structures. As a result, some of the moons become more homogeneous and lose their bilobate nature, while it is retained or even emphasised for other cases. The structurally fragile LABs sometimes undergo catastrophic tidal disruptions, stripping them of more than half their mass, leaving behind an oblate spheroid similar in shape to Dimorphos. For the fiducial merger case, we observe that the severity of the mass-loss occurs in discrete regimes of distance, where the innermost lobe is always fully disrupted closer in, while only ${\sim}10\%$ of the mass is lost in the next regime. Here, the tidal pull is generally directed along the longest principal axis of the body, leading to an accentuated neck between its two lobes. We believe this mechanism to be a potential explanation for the possible contact binary shape of Selam. Further out, the entire structure stays intact and any deformation of the body depends on its initially non-synchronous spin-orbit state. Employing a higher-resolution SFD, more true to the observed boulder diameters on rubble-pile asteroid surfaces, leads to more initial deformation from the merger, as having more small boulders in the aggregate smooths out contacts between the larger constituents. Nevertheless, the initial orientation of the most massive moonlet still highly affects the final shape of the merged object, and we observe similarities between the low- and high-resolution outcomes.

With these results in mind, along with the initially elongated shapes of the largest moonlets born in a binary asteroid formation scenario, we argue that it is necessary to take the non-sphericity of moonlets into account when modelling mergers in asteroid debris disks. Moreover, the use of irregularly shaped particles and polydisperse SFDs reproduces key granular mechanics crucial when capturing dynamical effects such as tidal distortion and disruption. Not only do these processes alter the shape of a rubble-pile satellite, but the timing of a tidal disruption event will also highly affect its resulting orbit, determining how far it migrates outwards and its spin-orbit state. As shown in this work, even a shift of 1\% in distance from the primary can lead to significant changes in the structural evolution of these objects. Furthermore, the change in dynamical behaviour for a merged satellite in the presence of a debris disk and the potential subsequent accretion of mass adds another layer of complexity to this problem that must be studied to properly understand the origin of atypical asteroid satellite shapes. Even so, with more data on the internal structure, particle size distribution and bulk density of Dimorphos from the Hera mission \citep{Michel_et_al_2022} and future missions involving in-situ observations of asteroid satellites such as Selam, along with more in-depth numerical studies of the debris disk model and tidal evolution of rubble-pile satellites, we will be able to further constrain their formation scenarios and dynamical history to get a more satisfying answer to how these fascinating objects came to be. 

\begin{acknowledgements}

J.W. and F.F. acknowledge funding from the Swiss National Science Foundation (SNSF) Ambizione grant No.\,193346. M.J. acknowledges support from SNSF project No.\,200021\_207359. We extend our gratitude to Harrison F. Agrusa, Sabina D. Raducan, Po-Yen Liu and Richard Cannon for helpful discussions and input throughout this project.

\end{acknowledgements}

\bibliographystyle{aa.bst}

\bibliography{References}

\begin{thebibliography}{71}
\expandafter\ifx\csname natexlab\endcsname\relax\def\natexlab#1{#1}\fi

\bibitem[{{Agrusa} {et~al.}(2022){Agrusa}, {Ferrari}, {Zhang}, {Richardson}, \&
  {Michel}}]{Agrusa_et_al_2022}
{Agrusa}, H.~F., {Ferrari}, F., {Zhang}, Y., {Richardson}, D.~C., \& {Michel},
  P. 2022, \psj, 3, 158

\bibitem[{{Agrusa} {et~al.}(2024){Agrusa}, {Zhang}, {Richardson}, {Pravec},
  {{\'C}uk}, {Michel}, {Ballouz}, {Jacobson}, {Scheeres}, {Walsh}, {Barnouin},
  {Daly}, {Palmer}, {Pajola}, {Lucchetti}, {Tusberti}, {DeMartini}, {Ferrari},
  {Meyer}, {Raducan}, \& {S{\'a}nchez}}]{Agrusa_et_al_2024}
{Agrusa}, H.~F., {Zhang}, Y., {Richardson}, D.~C., {et~al.} 2024, \psj, 5, 54

\bibitem[{Alliez {et~al.}(2022)Alliez, Cohen-Steiner, Hemmer, Portaneri, \&
  Rouxel-Labb{\'e}}]{cgal:alpha_wrap_3}
Alliez, P., Cohen-Steiner, D., Hemmer, M., Portaneri, C., \& Rouxel-Labb{\'e},
  M. 2022, in {CGAL} User and Reference Manual, {5.5} edn. ({CGAL Editorial
  Board})

\bibitem[{{Asphaug} \& {Benz}(1994)}]{Asphaug_&_Benz_1994}
{Asphaug}, E. \& {Benz}, W. 1994, \nat, 370, 120

\bibitem[{{Asphaug} \& {Benz}(1996)}]{Asphaug_&_Benz_1996}
{Asphaug}, E. \& {Benz}, W. 1996, \icarus, 121, 225

\bibitem[{{Barnouin} {et~al.}(2024){Barnouin}, {Ballouz}, {Marchi}, {Vincent},
  {Agrusa}, {Zhang}, {Ernst}, {Pajola}, {Tusberti}, {Lucchetti}, {Daly},
  {Palmer}, {Walsh}, {Michel}, {Sunshine}, {Rizos}, {Farnham}, {Richardson},
  {Parro}, {Murdoch}, {Robin}, {Hirabayashi}, {Kahout}, {Asphaug}, {Raducan},
  {Jutzi}, {Ferrari}, {Hasselmann}, {CampoBagatin}, {Chabot}, {Li}, {Cheng},
  {Nolan}, {Stickle}, {Karatekin}, {Dotto}, {Della Corte}, {Mazzotta Epifani},
  {Rossi}, {Gai}, {Deshapriya}, {Bertini}, {Zinzi}, {Trigo-Rodriguez},
  {Beccarelli}, {Ivanovski}, {Brucato}, {Poggiali}, {Zanotti}, {Amoroso},
  {Capannolo}, {Cremonese}, {Dall'Ora}, {Ieva}, {Impresario}, {Lavagn},
  {Modenini}, {Palumbo}, {Perna}, {Pirrotta}, {Tortora}, {Zannoni}, \&
  {Rivkin}}]{Barnouin_et_al_2024}
{Barnouin}, O., {Ballouz}, R.-L., {Marchi}, S., {et~al.} 2024, Nature
  Communications, 15, 6202

\bibitem[{{Barnouin} {et~al.}(2019){Barnouin}, {Daly}, {Palmer}, {Gaskell},
  {Weirich}, {Johnson}, {Al Asad}, {Roberts}, {Perry}, {Susorney}, {Daly},
  {Bierhaus}, {Seabrook}, {Espiritu}, {Nair}, {Nguyen}, {Neumann}, {Ernst},
  {Boynton}, {Nolan}, {Adam}, {Moreau}, {Rizk}, {Drouet D'Aubigny}, {Jawin},
  {Walsh}, {Michel}, {Schwartz}, {Ballouz}, {Mazarico}, {Scheeres}, {McMahon},
  {Bottke}, {Sugita}, {Hirata}, {Hirata}, {Watanabe}, {Burke}, {Dellagiustina},
  {Bennett}, {Lauretta}, {Osiris-Rex Team}, {Highsmith}, {Small},
  {Vokrouhlick{\'y}}, {Bowles}, {Brown}, {Donaldson Hanna}, {Warren}, {Brunet},
  {Chicoine}, {Desjardins}, {Gaudreau}, {Haltigin}, {Millington-Veloza},
  {Rubi}, {Aponte}, {Gorius}, {Lunsford}, {Allen}, {Grindlay}, {Guevel},
  {Hoak}, {Hong}, {Schrader}, {Bayron}, {Golubov}, {S{\'a}nchez}, {Stromberg},
  {Hirabayashi}, {Hartzell}, {Oliver}, {Rascon}, {Harch}, {Joseph}, {Squyres},
  {Richardson}, {Emery}, {McGraw}, {Ghent}, {Binzel}, {Asad}, {Johnson},
  {Philpott}, {Susorney}, {Cloutis}, {Hanna}, {Connolly}, {Ciceri},
  {Hildebrand}, {Ibrahim}, {Breitenfeld}, {Glotch}, {Rogers}, {Clark},
  {Ferrone}, {Thomas}, {Campins}, {Fernandez}, {Chang}, {Cheuvront}, {Trang},
  {Tachibana}, {Yurimoto}, {Brucato}, {Poggiali}, {Pajola}, {Dotto}, {Epifani},
  {Crombie}, {Lantz}, {Izawa}, {de Leon}, {Licandro}, {Garcia}, {Clemett},
  {Thomas-Keprta}, {van Wal}, {Yoshikawa}, {Bellerose}, {Bhaskaran}, {Boyles},
  {Chesley}, {Elder}, {Farnocchia}, {Harbison}, {Kennedy}, {Knight},
  {Martinez-Vlasoff}, {Mastrodemos}, {McElrath}, {Owen}, {Park}, {Rush},
  {Swanson}, {Takahashi}, {Velez}, {Yetter}, {Thayer}, {Adam}, {Antreasian},
  {Bauman}, {Bryan}, {Carcich}, {Corvin}, {Geeraert}, {Hoffman}, {Leonard},
  {Lessac-Chenen}, {Levine}, {McAdams}, {McCarthy}, {Nelson}, {Page},
  {Pelgrift}, {Sahr}, {Stakkestad}, {Stanbridge}, {Wibben}, {Williams},
  {Williams}, {Wolff}, {Hayne}, {Kubitschek}, {Barucci}, {Deshapriya},
  {Fornasier}, {Fulchignoni}, {Hasselmann}, {Merlin}, {Praet}, {Bierhaus},
  {Billett}, {Boggs}, {Buck}, {Carlson-Kelly}, {Cerna}, {Chaffin}, {Church},
  {Coltrin}, {Daly}, {Deguzman}, {Dubisher}, {Eckart}, {Ellis}, {Falkenstern},
  {Fisher}, {Fisher}, {Fleming}, {Fortney}, {Francis}, {Freund}, {Gonzales},
  {Haas}, {Hasten}, {Hauf}, {Hilbert}, {Howell}, {Jaen}, {Jayakody}, {Jenkins},
  {Johnson}, {Lefevre}, {Ma}, {Mario}, {Martin}, {May}, {McGee}, {Miller},
  {Miller}, {Miller}, {Mirfakhrai}, {Muhle}, {Norman}, {Olds}, {Parish},
  {Ryle}, {Schmitzer}, {Sherman}, {Skeen}, {Susak}, {Sutter}, {Tran}, {Welch},
  {Witherspoon}, {Wood}, {Zareski}, {Arvizu-Jakubicki}, {Asphaug}, {Audi},
  {Ballouz}, {Bandrowski}, {Becker}, {Becker}, {Bendall}, {Bennett},
  {Bloomenthal}, {Blum}, {Boynton}, {Brodbeck}, {Burke}, {Chojnacki}, {Colpo},
  {Contreras}, {Cutts}, {D'Aubigny}, {Dean}, {Dellagiustina}, {Diallo},
  {Drinnon}, {Drozd}, {Enos}, {Enos}, {Fellows}, {Ferro}, {Fisher},
  {Fitzgibbon}, {Fitzgibbon}, {Forelli}, {Forrester}, {Galinsky}, {Garcia},
  {Gardner}, {Golish}, {Habib}, {Hamara}, {Hammond}, {Hanley}, {Harshman},
  {Hergenrother}, {Herzog}, {Hill}, {Hoekenga}, {Hooven}, {Howell}, {Huettner},
  {Janakus}, {Jones}, {Kareta}, {Kidd}, {Kingsbury}, {Balram-Knutson},
  {Koelbel}, {Kreiner}, {Lambert}, {Lauretta}, {Lewin}, {Lovelace},
  {Loveridge}, {Lujan}, {Maleszewski}, {Malhotra}, {Marchese}, {McDonough},
  {Mogk}, {Morrison}, {Morton}, {Munoz}, {Nelson}, {Nolan}, {Padilla},
  {Pennington}, {Polit}, {Ramos}, {Reddy}, {Riehl}, {Rizk}, {Roper}, {Salazar},
  {Schwartz}, {Selznick}, {Shultz}, {Smith}, {Stewart}, {Sutton}, {Swindle},
  {Tang}, {Westermann}, {Wolner}, {Worden}, {Zega}, {Zeszut}, {Bjurstrom},
  {Bloomquist}, {Dickinson}, {Keates}, {Liang}, {Nifo}, {Taylor}, {Teti},
  {Caplinger}, {Bowles}, {Carter}, {Dickenshied}, {Doerres}, {Fisher}, {Hagee},
  {Hill}, {Miner}, {Noss}, {Piacentine}, {Smith}, {Toland}, {Wren}, {Bernacki},
  {Munoz}, {Watanabe}, {Sandford}, {Aqueche}, {Ashman}, {Barker}, {Bartels},
  {Berry}, {Bos}, {Burns}, {Calloway}, {Carpenter}, {Castro}, {Cosentino},
  {Donaldson}, {Dworkin}, {Cook}, {Emr}, {Everett}, {Fennell}, {Fleshman},
  {Folta}, {Gallagher}, {Garvin}, {Getzandanner}, {Glavin}, {Hull}, {Hyde},
  {Ido}, {Ingegneri}, {Jones}, {Kaotira}, {Lim}, {Liounis}, {Lorentson},
  {Lorenz}, {Lyzhoft}, {Mazarico}, {Mink}, {Moore}, {Moreau}, {Mullen}, {Nagy},
  {Neumann}, {Nuth}, {Poland}, {Reuter}, {Rhoads}, {Rieger}, {Rowlands},
  {Sallitt}, {Scroggins}, {Shaw}, {Simon}, {Swenson}, {Vasudeva}, {Wasser},
  {Zellar}, {Grossman}, {Johnston}, {Morris}, {Wendel}, {Burton}, {Keller},
  {McNamara}, {Messenger}, {Nakamura-Messenger}, {Nguyen}, {Righter}, {Queen},
  {Bellamy}, {Dill}, {Gardner}, {Giuntini}, {Key}, {Kissell}, {Patterson},
  {Vaughan}, {Wright}, {Gaskell}, {Le Corre}, {Li}, {Molaro}, {Palmer},
  {Siegler}, {Tricarico}, {Weirich}, {Zou}, {Ireland}, {Tait}, {Bland},
  {Anwar}, {Bojorquez-Murphy}, {Christensen}, {Haberle}, {Mehall}, {Rios},
  {Franchi}, {Rozitis}, {Beddingfield}, {Marshall}, {Brack}, {French},
  {McMahon}, {Scheeres}, {Jawin}, {McCoy}, {Russell}, {Killgore}, {Bottke},
  {Hamilton}, {Kaplan}, {Walsh}, {Bandfield}, {Clark}, {Chodas}, {Lambert},
  {Masterson}, {Daly}, {Freemantle}, {Seabrook}, {Barnouin}, {Craft}, {Daly},
  {Ernst}, {Espiritu}, {Holdridge}, {Jones}, {Nair}, {Nguyen}, {Peachey},
  {Perry}, {Plescia}, {Roberts}, {Steele}, {Turner}, {Backer}, {Edmundson},
  {Mapel}, {Milazzo}, {Sides}, {Manzoni}, {May}, {Delbo'}, {Libourel},
  {Michel}, {Ryan}, {Thuillet}, \& {Marty}}]{Barnouin_et_al_2019}
{Barnouin}, O.~S., {Daly}, M.~G., {Palmer}, E.~E., {et~al.} 2019, Nature
  Geoscience, 12, 247

\bibitem[{{Bottke} {et~al.}(1999){Bottke}, {Richardson}, {Michel}, \&
  {Love}}]{Bottke_et_al_1999}
{Bottke}, W.~F., J., {Richardson}, D.~C., {Michel}, P., \& {Love}, S.~G. 1999,
  \aj, 117, 1921

\bibitem[{{Brozovi{\'c}} {et~al.}(2018){Brozovi{\'c}}, {Benner}, {McMichael},
  {Giorgini}, {Pravec}, {Scheirich}, {Magri}, {Busch}, {Jao}, {Lee},
  {Snedeker}, {Silva}, {Slade}, {Semenov}, {Nolan}, {Taylor}, {Howell}, \&
  {Lawrence}}]{Brozovic_et_al_2018}
{Brozovi{\'c}}, M., {Benner}, L. A.~M., {McMichael}, J.~G., {et~al.} 2018,
  \icarus, 300, 115

\bibitem[{{Brozovi{\'c}} {et~al.}(2022){Brozovi{\'c}}, {Benner}, {Naidu},
  {Busch}, {Giorgini}, {Lazio}, \& {Hall}}]{Brozovic_et_al_2022}
{Brozovi{\'c}}, M., {Benner}, L.~A.~M., {Naidu}, S.~P., {et~al.} 2022, in LPI
  Contributions, Vol. 2681, Apophis T-7 Years: Knowledge Opportunities for the
  Science of Planetary Defense, 2023

\bibitem[{{Burnett} {et~al.}(2024){Burnett}, {Fodde}, \&
  {Ferrari}}]{Burnett_Fodde_&_Ferrari_2025}
{Burnett}, E.~R., {Fodde}, I., \& {Ferrari}, F. 2024, arXiv e-prints,
  arXiv:2410.09266

\bibitem[{{Chabot} {et~al.}(2024){Chabot}, {Rivkin}, {Cheng}, {Barnouin},
  {Fahnestock}, {Richardson}, {Stickle}, {Thomas}, {Ernst}, {Terik Daly},
  {Dotto}, {Zinzi}, {Chesley}, {Moskovitz}, {Barbee}, {Abell}, {Agrusa},
  {Bannister}, {Beccarelli}, {Bekker}, {Bruck Syal}, {Buratti}, {Busch}, {Campo
  Bagatin}, {Chatelain}, {Chocron}, {Collins}, {Conversi}, {Davison},
  {DeCoster}, {Prasanna Deshapriya}, {Eggl}, {Espiritu}, {Farnham}, {Ferrais},
  {Ferrari}, {F{\"o}hring}, {Fuentes-Mu{\~n}oz}, {Gai}, {Giordano}, {Glenar},
  {Gomez}, {Graninger}, {Green}, {Greenstreet}, {Hasselmann}, {Herreros},
  {Hirabayashi}, {Hus{\'a}rik}, {Ieva}, {Ivanovski}, {Jackson}, {Jehin},
  {Jutzi}, {Karatekin}, {Knight}, {Kolokolova}, {Kumamoto}, {K{\"u}ppers}, {La
  Forgia}, {Lazzarin}, {Li}, {Lister}, {Lolachi}, {Lucas}, {Lucchetti},
  {Luther}, {Makadia}, {Mazzotta Epifani}, {McMahon}, {Merisio}, {Merrill},
  {Meyer}, {Michel}, {Micheli}, {Migliorini}, {Minker}, {Modenini}, {Moreno},
  {Murdoch}, {Murphy}, {Naidu}, {Nair}, {Nakano}, {Opitom}, {Orm{\"o}},
  {Michael Owen}, {Pajola}, {Palmer}, {Palumbo}, {Panicucci}, {Parro}, {Pearl},
  {Penttil{\"a}}, {Perna}, {Petrescu}, {Pravec}, {Raducan}, {Ramesh},
  {Ridden-Harper}, {Rizos}, {Rossi}, {Roth}, {Ro{\.z}ek}, {Rozitis}, {Ryan},
  {Ryan}, {S{\'a}nchez}, {Santana-Ros}, {Scheeres}, {Scheirich}, {Senel},
  {Snodgrass}, {Soldini}, {Souami}, {Statler}, {Street}, {Stubbs}, {Sunshine},
  {Tan}, {Tancredi}, {Tinsman}, {Tortora}, {Tusberti}, {Walker}, {Waller},
  {W{\"u}nnemann}, {Zannoni}, \& {Zhang}}]{Chabot_et_al_2024}
{Chabot}, N.~L., {Rivkin}, A.~S., {Cheng}, A.~F., {et~al.} 2024, \psj, 5, 49

\bibitem[{Da {et~al.}(2023)Da, Loriot, \& Yvinec}]{cgal:alpha_shapes_3}
Da, T. K.~F., Loriot, S., \& Yvinec, M. 2023, in {CGAL} User and Reference
  Manual, {5.5.2} edn. ({CGAL Editorial Board})

\bibitem[{{Daly} {et~al.}(2023){Daly}, {Ernst}, {Barnouin}, {Chabot}, {Rivkin},
  {Cheng}, {Adams}, {Agrusa}, {Abel}, {Alford}, {Asphaug}, {Atchison},
  {Badger}, {Baki}, {Ballouz}, {Bekker}, {Bellerose}, {Bhaskaran}, {Buratti},
  {Cambioni}, {Chen}, {Chesley}, {Chiu}, {Collins}, {Cox}, {DeCoster},
  {Ericksen}, {Espiritu}, {Faber}, {Farnham}, {Ferrari}, {Fletcher}, {Gaskell},
  {Graninger}, {Haque}, {Harrington-Duff}, {Hefter}, {Herreros}, {Hirabayashi},
  {Huang}, {Hsieh}, {Jacobson}, {Jenkins}, {Jensenius}, {John}, {Jutzi},
  {Kohout}, {Krueger}, {Laipert}, {Lopez}, {Luther}, {Lucchetti}, {Mages},
  {Marchi}, {Martin}, {McQuaide}, {Michel}, {Moskovitz}, {Murphy}, {Murdoch},
  {Naidu}, {Nair}, {Nolan}, {Orm{\"o}}, {Pajola}, {Palmer}, {Peachey},
  {Pravec}, {Raducan}, {Ramesh}, {Ramirez}, {Reynolds}, {Richman}, {Robin},
  {Rodriguez}, {Roufberg}, {Rush}, {Sawyer}, {Scheeres}, {Scheirich},
  {Schwartz}, {Shannon}, {Shapiro}, {Shearer}, {Smith}, {Steele}, {Steckloff},
  {Stickle}, {Sunshine}, {Superfin}, {Tarzi}, {Thomas}, {Thomas},
  {Trigo-Rodr{\'\i}guez}, {Tropf}, {Vaughan}, {Velez}, {Waller}, {Wilson},
  {Wortman}, \& {Zhang}}]{Daly_et_al_2023a}
{Daly}, R.~T., {Ernst}, C.~M., {Barnouin}, O.~S., {et~al.} 2023, \nat, 616, 443

\bibitem[{{Daly} {et~al.}(2024){Daly}, {Ernst}, {Barnouin}, {Gaskell}, {Nair},
  {Agrusa}, {Chabot}, {Cheng}, {Dotto}, {Mazzotta Epifani}, {Espiritu},
  {Farnham}, {Palmer}, {Pravec}, {Rivkin}, {Waller}, {Zinzi}, {DART Team}, \&
  {LICIACube Team}}]{Daly_et_al_2024}
{Daly}, R.~T., {Ernst}, C.~M., {Barnouin}, O.~S., {et~al.} 2024, \psj, 5, 24

\bibitem[{Edelsbrunner \& M{\"u}cke(1994)}]{Edelsbrunner_&_Mucke_1994}
Edelsbrunner, H. \& M{\"u}cke, E.~P. 1994, ACM Transactions On Graphics (TOG),
  13, 43

\bibitem[{{Ferrari} {et~al.}(2020){Ferrari}, {Lavagna}, \&
  {Blazquez}}]{Ferrari_et_al_2020}
{Ferrari}, F., {Lavagna}, M., \& {Blazquez}, E. 2020, \mnras, 492, 749

\bibitem[{{Ferrari} {et~al.}(2022){Ferrari}, {Raducan}, {Soldini}, \&
  {Jutzi}}]{Ferrari_et_al_2022}
{Ferrari}, F., {Raducan}, S.~D., {Soldini}, S., \& {Jutzi}, M. 2022, \psj, 3,
  177

\bibitem[{{Ferrari} \& {Tanga}(2020)}]{Ferrari_&_Tanga_2020}
{Ferrari}, F. \& {Tanga}, P. 2020, \icarus, 350, 113871

\bibitem[{{Ferrari} \& {Tanga}(2022)}]{Ferrari_&_Tanga_2022}
{Ferrari}, F. \& {Tanga}, P. 2022, \icarus, 378, 114914

\bibitem[{{Ferrari} {et~al.}(2017){Ferrari}, {Tasora}, {Masarati}, \&
  {Lavagna}}]{Ferrari_et_al_2017}
{Ferrari}, F., {Tasora}, A., {Masarati}, P., \& {Lavagna}, M. 2017, Multibody
  System Dynamics, 39, 3

\bibitem[{{Flynn} {et~al.}(2018){Flynn}, {Consolmagno}, {Brown}, \&
  {Macke}}]{Flynn_et_al_2018}
{Flynn}, G.~J., {Consolmagno}, G.~J., {Brown}, P., \& {Macke}, R.~J. 2018,
  Chemie der Erde / Geochemistry, 78, 269

\bibitem[{{Fujiwara} {et~al.}(2006){Fujiwara}, {Kawaguchi}, {Yeomans}, {Abe},
  {Mukai}, {Okada}, {Saito}, {Yano}, {Yoshikawa}, {Scheeres}, {Barnouin-Jha},
  {Cheng}, {Demura}, {Gaskell}, {Hirata}, {Ikeda}, {Kominato}, {Miyamoto},
  {Nakamura}, {Nakamura}, {Sasaki}, \& {Uesugi}}]{Fujiwara_et_al_2006}
{Fujiwara}, A., {Kawaguchi}, J., {Yeomans}, D.~K., {et~al.} 2006, Science, 312,
  1330

\bibitem[{{Holsapple} \& {Michel}(2006)}]{Holsapple_&_Michel_2006}
{Holsapple}, K.~A. \& {Michel}, P. 2006, \icarus, 183, 331

\bibitem[{{Holsapple} \& {Michel}(2008)}]{Holsapple_&_Michel_2008}
{Holsapple}, K.~A. \& {Michel}, P. 2008, \icarus, 193, 283

\bibitem[{{Jacobson} \& {Scheeres}(2011)}]{Jacobson_&_Scheeres_2011a}
{Jacobson}, S.~A. \& {Scheeres}, D.~J. 2011, \icarus, 214, 161

\bibitem[{{Jacobson} {et~al.}(2014){Jacobson}, {Scheeres}, \&
  {McMahon}}]{Jacobson_et_al_2014}
{Jacobson}, S.~A., {Scheeres}, D.~J., \& {McMahon}, J. 2014, \apj, 780, 60

\bibitem[{{Jutzi} \& {Asphaug}(2015)}]{Jutzi_&_Asphaug_2015}
{Jutzi}, M. \& {Asphaug}, E. 2015, Science, 348, 1355

\bibitem[{{Jutzi} \& {Benz}(2017)}]{Jutzi_&_Benz_2017}
{Jutzi}, M. \& {Benz}, W. 2017, \aap, 597, A62

\bibitem[{{Korycansky} \& {Asphaug}(2006)}]{Korycansky_&_Asphaug_2006}
{Korycansky}, D.~G. \& {Asphaug}, E. 2006, \icarus, 181, 605

\bibitem[{{Korycansky} \& {Asphaug}(2009)}]{Korycansky_&_Asphaug_2009}
{Korycansky}, D.~G. \& {Asphaug}, E. 2009, \icarus, 204, 316

\bibitem[{{Leleu} {et~al.}(2018){Leleu}, {Jutzi}, \&
  {Rubin}}]{Leleu_et_al_2018}
{Leleu}, A., {Jutzi}, M., \& {Rubin}, M. 2018, Nature Astronomy, 2, 555

\bibitem[{{Levison} {et~al.}(2024){Levison}, {Marchi}, {Noll}, {Spencer},
  {Statler}, {Bell}, {Bierhaus}, {Binzel}, {Bottke}, {Britt}, {Brown}, {Buie},
  {Christensen}, {Dello Russo}, {Emery}, {Grundy}, {Hahn}, {Hamilton},
  {Howett}, {Kaplan}, {Kretke}, {Lauer}, {Manzoni}, {Marschall}, {Martin},
  {May}, {Mottola}, {Olkin}, {P{\"a}tzold}, {Parker}, {Porter}, {Preusker},
  {Protopapa}, {Reuter}, {Robbins}, {Salmon}, {Simon}, {Stern}, {Sunshine},
  {Wong}, {Weaver}, {Adam}, {Ancheta}, {Andrews}, {Anwar}, {Barnouin},
  {Beasley}, {Berry}, {Birath}, {Bolin}, {Booco}, {Burns}, {Campbell},
  {Carpenter}, {Crombie}, {Effertz}, {Eifert}, {Ellis}, {Faiks}, {Fischetti},
  {Fleming}, {Francis}, {Franco}, {Freund}, {Gallagher}, {Geeraert}, {Gobat},
  {Gorgas}, {Granat}, {Gray}, {Haas}, {Harch}, {Hegedus}, {Isabelle},
  {Jackson}, {Jacob}, {Jennings}, {Kaufmann}, {Keeney}, {Kennedy}, {Lauffer},
  {Lessac-Chenen}, {Leonard}, {Levine}, {Lunsford}, {Martin}, {McAdams},
  {Mehall}, {Merkley}, {Miller}, {Montanaro}, {Montgomery}, {Murphy}, {Myers},
  {Nelson}, {Ocampo}, {Olds}, {Pelgrift}, {Perkins}, {Pineau}, {Poland},
  {Ramanan}, {Rose}, {Sahr}, {Short}, {Solanki}, {Stanbridge}, {Sutter},
  {Talpas}, {Taylor}, {Treiu}, {Vermeer}, {Vincent}, {Wallace}, {Weigle},
  {Wibben}, {Wiens}, {Wilson}, \& {Zhao}}]{Levison_et_al_2024}
{Levison}, H.~F., {Marchi}, S., {Noll}, K.~S., {et~al.} 2024, \nat, 629, 1015

\bibitem[{{Levison} {et~al.}(2021){Levison}, {Olkin}, {Noll}, {Marchi}, {Bell},
  {Bierhaus}, {Binzel}, {Bottke}, {Britt}, {Brown}, {Buie}, {Christensen},
  {Emery}, {Grundy}, {Hamilton}, {Howett}, {Mottola}, {P{\"a}tzold}, {Reuter},
  {Spencer}, {Statler}, {Stern}, {Sunshine}, {Weaver}, \&
  {Wong}}]{Levison_et_al_2021}
{Levison}, H.~F., {Olkin}, C.~B., {Noll}, K.~S., {et~al.} 2021, \psj, 2, 171

\bibitem[{{Madeira} \& {Charnoz}(2024)}]{Madeira_&_Charnoz_2024}
{Madeira}, G. \& {Charnoz}, S. 2024, \icarus, 409, 115871

\bibitem[{{Marohnic} {et~al.}(2023){Marohnic}, {DeMartini}, {Richardson},
  {Zhang}, \& {Walsh}}]{Marohnic_et_al_2023}
{Marohnic}, J.~C., {DeMartini}, J.~V., {Richardson}, D.~C., {Zhang}, Y., \&
  {Walsh}, K.~J. 2023, \psj, 4, 245

\bibitem[{{Marohnic} {et~al.}(2021){Marohnic}, {Richardson}, {McKinnon},
  {Agrusa}, {DeMartini}, {Cheng}, {Stern}, {Olkin}, {Weaver}, {Spencer}, \&
  {New Horizons Science Team}}]{Mahronic_et_al_2021}
{Marohnic}, J.~C., {Richardson}, D.~C., {McKinnon}, W.~B., {et~al.} 2021,
  \icarus, 356, 113824

\bibitem[{{Marti} {et~al.}(2024){Marti}, {Quinteros}, {Mikesell}, {Margerin},
  {Delage}, \& {Murdoch}}]{Marti_et_al_2024}
{Marti}, J., {Quinteros}, S., {Mikesell}, D., {et~al.} 2024, in European
  Planetary Science Congress, EPSC2024--488

\bibitem[{{Merrill} {et~al.}(2024){Merrill}, {Kubas}, {Meyer}, \&
  {Raducan}}]{Merrill_et_al_2024}
{Merrill}, C.~C., {Kubas}, A.~R., {Meyer}, A.~J., \& {Raducan}, S.~D. 2024,
  \aap, 684, L20

\bibitem[{{Michel} {et~al.}(2022){Michel}, {K{\"u}ppers}, {Bagatin}, {Carry},
  {Charnoz}, {de Leon}, {Fitzsimmons}, {Gordo}, {Green}, {H{\'e}rique}, {Juzi},
  {Karatekin}, {Kohout}, {Lazzarin}, {Murdoch}, {Okada}, {Palomba}, {Pravec},
  {Snodgrass}, {Tortora}, {Tsiganis}, {Ulamec}, {Vincent}, {W{\"u}nnemann},
  {Zhang}, {Raducan}, {Dotto}, {Chabot}, {Cheng}, {Rivkin}, {Barnouin},
  {Ernst}, {Stickle}, {Richardson}, {Thomas}, {Arakawa}, {Miyamoto},
  {Nakamura}, {Sugita}, {Yoshikawa}, {Abell}, {Asphaug}, {Ballouz}, {Bottke},
  {Lauretta}, {Walsh}, {Martino}, \& {Carnelli}}]{Michel_et_al_2022}
{Michel}, P., {K{\"u}ppers}, M., {Bagatin}, A.~C., {et~al.} 2022, \psj, 3, 160

\bibitem[{{Movshovitz} {et~al.}(2012){Movshovitz}, {Asphaug}, \&
  {Korycansky}}]{Movshovitz_et_al_2012}
{Movshovitz}, N., {Asphaug}, E., \& {Korycansky}, D. 2012, \apj, 759, 93

\bibitem[{{Naidu} {et~al.}(2024){Naidu}, {Chesley}, {Moskovitz}, {Thomas},
  {Meyer}, {Pravec}, {Scheirich}, {Farnocchia}, {Scheeres}, {Brozovic},
  {Benner}, {Rivkin}, \& {Chabot}}]{Naidu_et_al_2024}
{Naidu}, S.~P., {Chesley}, S.~R., {Moskovitz}, N., {et~al.} 2024, \psj, 5, 74

\bibitem[{{Pajola} {et~al.}(2024){Pajola}, {Tusberti}, {Lucchetti}, {Barnouin},
  {Cambioni}, {Ernst}, {Dotto}, {Daly}, {Poggiali}, {Hirabayashi}, {Nakano},
  {Epifani}, {Chabot}, {Della Corte}, {Rivkin}, {Agrusa}, {Zhang}, {Penasa},
  {Ballouz}, {Ivanovski}, {Murdoch}, {Rossi}, {Robin}, {Ieva}, {Vincent},
  {Ferrari}, {Raducan}, {Campo-Bagatin}, {Parro}, {Benavidez}, {Tancredi},
  {Karatekin}, {Trigo-Rodriguez}, {Sunshine}, {Farnham}, {Asphaug},
  {Deshapriya}, {Hasselmann}, {Beccarelli}, {Schwartz}, {Abell}, {Michel},
  {Cheng}, {Brucato}, {Zinzi}, {Amoroso}, {Pirrotta}, {Impresario}, {Bertini},
  {Capannolo}, {Caporali}, {Ceresoli}, {Cremonese}, {Dall'Ora}, {Gai},
  {Casajus}, {Gramigna}, {Manghi}, {Lavagna}, {Lombardo}, {Modenini},
  {Palumbo}, {Perna}, {Tortora}, {Zannoni}, \& {Zanotti}}]{Pajola_et_al_2024}
{Pajola}, M., {Tusberti}, F., {Lucchetti}, A., {et~al.} 2024, Nature
  Communications, 15, 6205

\bibitem[{{Porco} {et~al.}(2007){Porco}, {Thomas}, {Weiss}, \&
  {Richardson}}]{Porco_et_al_2007}
{Porco}, C.~C., {Thomas}, P.~C., {Weiss}, J.~W., \& {Richardson}, D.~C. 2007,
  Science, 318, 1602

\bibitem[{{Pravec} {et~al.}(2016){Pravec}, {Scheirich}, {Ku{\v{s}}nir{\'a}k},
  {Hornoch}, {Gal{\'a}d}, {Naidu}, {Pray}, {Vil{\'a}gi}, {Gajdo{\v{s}}},
  {Korno{\v{s}}}, {Krugly}, {Cooney}, {Gross}, {Terrell}, {Gaftonyuk},
  {Pollock}, {Hus{\'a}rik}, {Chiorny}, {Stephens}, {Durkee}, {Reddy}, {Dyvig},
  {Vra{\v{s}}til}, {{\v{Z}}i{\v{z}}ka}, {Mottola}, {Hellmich}, {Oey},
  {Benishek}, {Kryszczy{\'n}ska}, {Higgins}, {Ries}, {Marchis}, {Baek},
  {Macomber}, {Inasaridze}, {Kvaratskhelia}, {Ayvazian}, {Rumyantsev}, {Masi},
  {Colas}, {Lecacheux}, {Montaigut}, {Leroy}, {Brown}, {Krzeminski}, {Molotov},
  {Reichart}, {Haislip}, \& {LaCluyze}}]{Pravec_et_al_2016}
{Pravec}, P., {Scheirich}, P., {Ku{\v{s}}nir{\'a}k}, P., {et~al.} 2016,
  \icarus, 267, 267

\bibitem[{Raducan {et~al.}(2024)Raducan, Jutzi, Cheng, Zhang, Barnouin,
  Collins, Daly, Davison, Ernst, Farnham, Ferrari, Hirabayashi, Kumamoto,
  Michel, Murdoch, Nakano, Pajola, Rossi, Agrusa, Barbee, Syal, Chabot, Dotto,
  Fahnestock, Hasselmann, Herreros, Ivanovski, Li, Lucchetti, Luther, Ormö,
  Owen, Pravec, Rivkin, Robin, Sánchez, Tusberti, Wünnemann, Zinzi, Epifani,
  Manzoni, \& May}]{Raducan_et_al_2024a}
Raducan, S.~D., Jutzi, M., Cheng, A.~F., {et~al.} 2024, Nature Astronomy

\bibitem[{{Raducan} {et~al.}(2024){Raducan}, {Jutzi}, {Merrill}, {Michel},
  {Zhang}, {Hirabayashi}, \& {Mainzer}}]{Raducan_et_al_2024b}
{Raducan}, S.~D., {Jutzi}, M., {Merrill}, C.~C., {et~al.} 2024, \psj, 5, 79

\bibitem[{{Raducan} {et~al.}(2022){Raducan}, {Jutzi}, {Zhang}, {Orm{\"o}}, \&
  {Michel}}]{Raducan_et_al_2022}
{Raducan}, S.~D., {Jutzi}, M., {Zhang}, Y., {Orm{\"o}}, J., \& {Michel}, P.
  2022, \aap, 665, L10

\bibitem[{{Raducan} {et~al.}(2025){Raducan}, {Madeira}, {Agrusa}, {Merrill},
  {Marschall}, {Ferrari}, {Wimarsson}, {Charnoz}, \&
  {Jutzi}}]{Raducan_et_al_2025}
{Raducan}, S.~D., {Madeira}, G., {Agrusa}, H.~F., {et~al.} 2025, Nature
  Communications, in review

\bibitem[{{Richardson} {et~al.}(1998){Richardson}, {Bottke}, \&
  {Love}}]{Richardson_Bottke_&_Love_1998}
{Richardson}, D.~C., {Bottke}, W.~F., \& {Love}, S.~G. 1998, \icarus, 134, 47

\bibitem[{{Rivkin} {et~al.}(2021){Rivkin}, {Chabot}, {Stickle}, {Thomas},
  {Richardson}, {Barnouin}, {Fahnestock}, {Ernst}, {Cheng}, {Chesley}, {Naidu},
  {Statler}, {Barbee}, {Agrusa}, {Moskovitz}, {Terik Daly}, {Pravec},
  {Scheirich}, {Dotto}, {Della Corte}, {Michel}, {K{\"u}ppers}, {Atchison}, \&
  {Hirabayashi}}]{Rivkin_et_al_2021}
{Rivkin}, A.~S., {Chabot}, N.~L., {Stickle}, A.~M., {et~al.} 2021, \psj, 2, 173

\bibitem[{{Robin} {et~al.}(2024){Robin}, {Duchene}, {Murdoch}, {Vincent},
  {Lucchetti}, {Pajola}, {Ernst}, {Daly}, {Barnouin}, {Raducan}, {Michel},
  {Hirabayashi}, {Stott}, {Cuervo}, {Jawin}, {Trigo-Rodriguez}, {Parro},
  {Sunday}, {Vivet}, {Mimoun}, {Rivkin}, \& {Chabot}}]{Robin_et_al_2024}
{Robin}, C.~Q., {Duchene}, A., {Murdoch}, N., {et~al.} 2024, Nature
  Communications, 15, 6203

\bibitem[{Roche(1847)}]{Roche_1847}
Roche, E. 1847, Acad. Sci. Lett. Montpelier. Mem. Section Sci., 1, 243

\bibitem[{Rubincam(2000)}]{Rubincam_2000}
Rubincam, D.~P. 2000, Icarus (New York, N.Y. 1962), 148, 2

\bibitem[{{S{\'a}nchez} \& {Scheeres}(2011)}]{Sanchez_&_Scheeres_2011}
{S{\'a}nchez}, P. \& {Scheeres}, D.~J. 2011, \apj, 727, 120

\bibitem[{Schib {et~al.}(2025{\natexlab{a}})Schib, Mordasini, Emsenhuber, \&
  Helled}]{Schib_et_al_2025a}
Schib, O., Mordasini, C., Emsenhuber, A., \& Helled, R. 2025{\natexlab{a}},
  \aap~in press [\eprint[arXiv]{2510.02436}]

\bibitem[{Schib {et~al.}(2025{\natexlab{b}})Schib, Mordasini, Emsenhuber, \&
  Helled}]{Schib_et_al_2025b}
Schib, O., Mordasini, C., Emsenhuber, A., \& Helled, R. 2025{\natexlab{b}},
  \aap~in press [\eprint[arXiv]{2510.02437}]

\bibitem[{Schwartz {et~al.}(2012)Schwartz, Richardson, \&
  Michel}]{Schwartz_et_al_2012}
Schwartz, S.~R., Richardson, D.~C., \& Michel, P. 2012, Granular Matter, 14,
  363

\bibitem[{{Tardivel} {et~al.}(2018){Tardivel}, {S{\'a}nchez}, \&
  {Scheeres}}]{Tardivel_et_al_2018}
{Tardivel}, S., {S{\'a}nchez}, P., \& {Scheeres}, D.~J. 2018, \icarus, 304, 192

\bibitem[{Tasora(2017)}]{Chrono_integrators2017}
Tasora, A. 2017, {Time integration in Chrono::Engine}

\bibitem[{Tasora(2020)}]{Chrono_rotations2020}
Tasora, A. 2020, {Rotations in Chrono::Engine}

\bibitem[{Tasora {et~al.}(2016)Tasora, Serban, Mazhar, Pazouki, Melanz,
  Fleischmann, Taylor, Sugiyama, \& Negrut}]{Chrono2016}
Tasora, A., Serban, R., Mazhar, H., {et~al.} 2016, in High Performance
  Computing in Science and Engineering – Lecture Notes in Computer Science,
  ed. T.~Kozubek (Springer), 19--49

\bibitem[{{Tiscareno} {et~al.}(2013){Tiscareno}, {Hedman}, {Burns}, \&
  {Castillo-Rogez}}]{Tiscareno_et_al_2013}
{Tiscareno}, M.~S., {Hedman}, M.~M., {Burns}, J.~A., \& {Castillo-Rogez}, J.
  2013, \apjl, 765, L28

\bibitem[{{Walsh} \& {Richardson}(2006)}]{Walsh_&_Richardson_2006}
{Walsh}, K.~J. \& {Richardson}, D.~C. 2006, \icarus, 180, 201

\bibitem[{{Walsh} {et~al.}(2008){Walsh}, {Richardson}, \&
  {Michel}}]{Walsh_et_al_2008}
{Walsh}, K.~J., {Richardson}, D.~C., \& {Michel}, P. 2008, \nat, 454, 188

\bibitem[{{Walsh} {et~al.}(2012){Walsh}, {Richardson}, \&
  {Michel}}]{Walsh_et_al_2012}
{Walsh}, K.~J., {Richardson}, D.~C., \& {Michel}, P. 2012, \icarus, 220, 514

\bibitem[{{Wimarsson} {et~al.}(2024){Wimarsson}, {Xiang}, {Ferrari}, {Jutzi},
  {Madeira}, {Raducan}, \& {S{\'a}nchez}}]{Wimarsson_et_al_2024}
{Wimarsson}, J., {Xiang}, Z., {Ferrari}, F., {et~al.} 2024, \icarus, 421,
  116223

\bibitem[{{Zhang} \& {Lin}(2020)}]{Zhang_&_Lin_2020}
{Zhang}, Y. \& {Lin}, D. N.~C. 2020, Nature Astronomy, 4, 852

\bibitem[{{Zhang} \& {Michel}(2020)}]{Zhang_&_Michel_2020}
{Zhang}, Y. \& {Michel}, P. 2020, \aap, 640, A102

\bibitem[{{Zhang} {et~al.}(2021){Zhang}, {Michel}, {Richardson}, {Barnouin},
  {Agrusa}, {Tsiganis}, {Manzoni}, \& {May}}]{Zhang_et_al_2021}
{Zhang}, Y., {Michel}, P., {Richardson}, D.~C., {et~al.} 2021, \icarus, 362,
  114433

\bibitem[{{Zhang} {et~al.}(2017){Zhang}, {Richardson}, {Barnouin}, {Maurel},
  {Michel}, {Schwartz}, {Ballouz}, {Benner}, {Naidu}, \&
  {Li}}]{Zhang_et_al_2017}
{Zhang}, Y., {Richardson}, D.~C., {Barnouin}, O.~S., {et~al.} 2017, \icarus,
  294, 98

\end{thebibliography}

\begin{appendix}
\onecolumn

\section{Rotations and quaternions in \texttt{GRAINS}}\label{appendix:quaternions}

In \texttt{Chrono}, the orientation of each particle is tracked using unit quaternions, which are four-dimensional vectors that have the following form

\begin{equation}
\boldsymbol{q} = q_0+ q_1 \boldsymbol{i} + q_2\boldsymbol{j} + q_3 \boldsymbol{k},
\end{equation}

\noindent 
where $q_0,\ q_1,\ q_2,\ q_3$ are real numbers and $\boldsymbol{i},\ \boldsymbol{j},\ \boldsymbol{k}$ are basis elements fulfilling $i^2 = j^2 = k^2 = ijk = -1$. Moreover, they always satisfy $|\boldsymbol{q}| = 1$. As a result, each particle's rotational state can conveniently be represented by just four numbers, with $\boldsymbol{q} = \{1,0,0,0\}$ being the initial state. Another notation $\boldsymbol{q}=(s,\boldsymbol{u})$ is based on the fact that a quaternion consists of a scalar $s$ and an imaginary vector $\boldsymbol{u}$.

Assuming we want to rotate a given particle by an angle $\alpha$ around its centre-of-mass in an $x$, $y$, $z$ Cartesian coordinate system, this can be achieved using quaternion multiplication. This is possible due to the fact that the product of a quaternion $\boldsymbol{q}$ by its conjugate $\boldsymbol{q}^* = (q_0-q_1i-q_2j-q_3k)$, is a quaternion without its imaginary part as $\boldsymbol{qq}^* = (q_0^2+q_1^2+q_2^2+q_3^2)$. Hence, it follows that a quaternion describing the rotational state of particle $b$, $\boldsymbol{p}_b$, can be transformed by another quaternion, $\boldsymbol{q}$, representing some rotation. Given a rotation $\alpha$ around some general unit vector $\boldsymbol{n}$, this quaternion has the form $\boldsymbol{q}=\{\cos{(\alpha/2)}, n_x\sin{(\alpha/2)}, n_y\sin{(\alpha/2)}, n_z\sin{(\alpha/2)}\}$. We then carry out the operation to obtain a new, transformed quaternion 

\begin{equation}\label{eq:quaternion_rotation}
    \boldsymbol{p}_b' = \boldsymbol{q} \boldsymbol{p}_b \boldsymbol{q}^*.
\end{equation}

Since we are using unit quaternions, the norm of $\boldsymbol{p}_b$ remains unchanged. Furthermore, this treatment can also be applied to vectors representing position, velocity and angular velocity, which follows from the fact that some vector $\boldsymbol{v}$ can be represented by the quaternion $\boldsymbol{p}_v = (0,\boldsymbol{v})$. Applying Eq. (\ref{eq:quaternion_rotation}) then yields the transformed entity $\boldsymbol{p}_v'=(0,\boldsymbol{v}')$. For more details regarding quaternion properties and their use in \texttt{Chrono}, see \citet{Chrono_rotations2020}.

\section{Bilobate DEEVE fitting}\label{appendix:DEEVE_fits}

In order to determine the DEEVE of each lobe after the merger and estimate their semi-major axes, we implemented a simple algorithm. First, we assumed that each particle belonged to the same parent moonlet pre- and post-merger and computed the barycenter of each lobe, $\boldsymbol{s}_A$ and $\boldsymbol{s}_B$. Second, we iterated over each particle and evaluated their position relative each barycenter regardless of which lobe is their initial parent to account for mixing. For a given particle $i$, if the distance $d_A = |\boldsymbol{r}_i - \boldsymbol{s}_A|$ is less than $d_B = |\boldsymbol{r}_i - \boldsymbol{s}_B|$, it belonged to lobe A after the merger and vice versa. After repeating this process for all particles, we calculated the new barycenter and total moment of inertia for each lobe. Given the soft impact and the little-to-no mixing, one iteration of this algorithm was enough to give good visual DEEVE fits.

\section{Complementary plots}\label{appendix:plots}

\begin{figure*}[!h]
    \resizebox{\hsize}{!}{\includegraphics{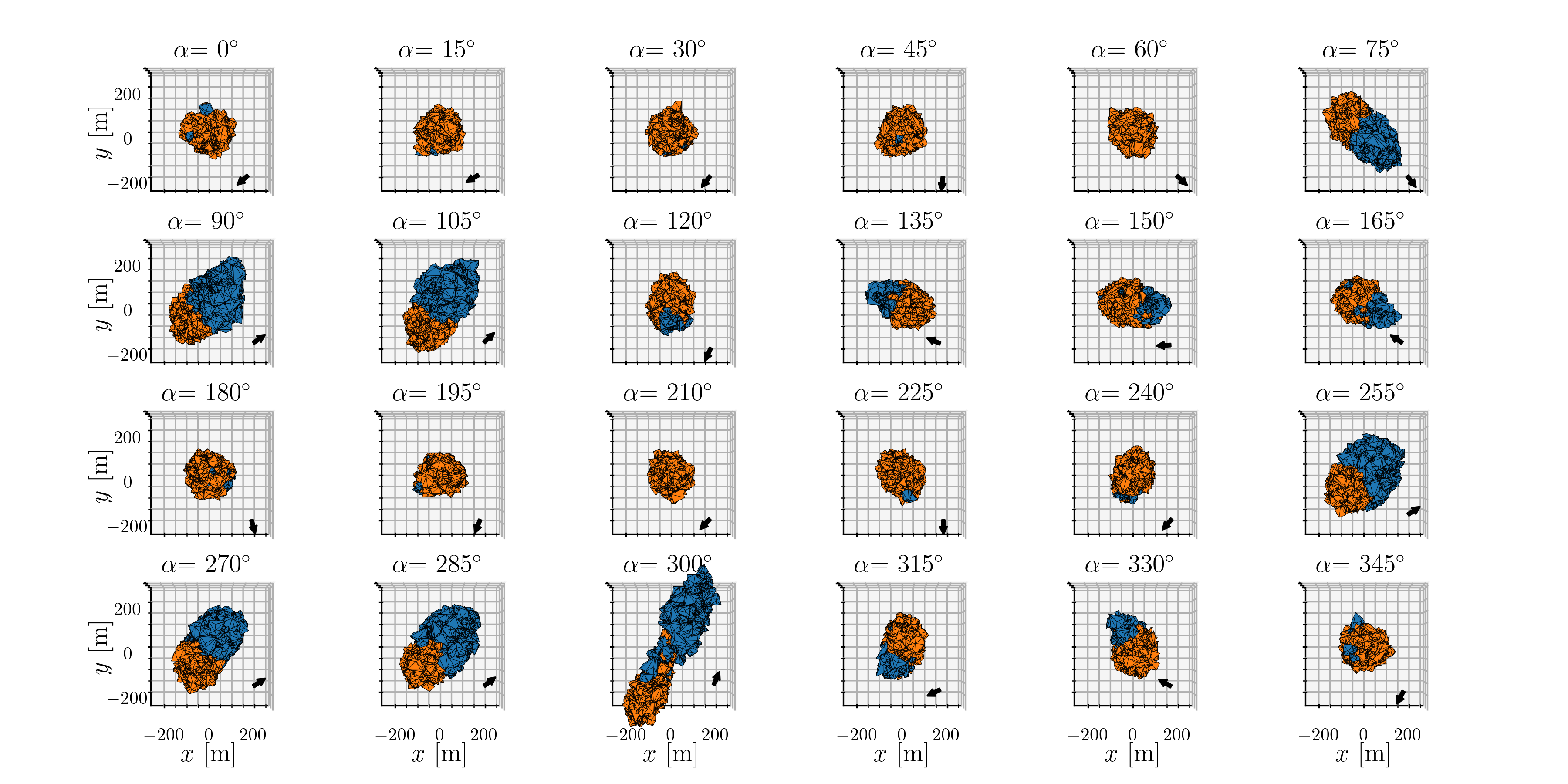}}
    \caption{The shapes of the bodies in Fig.~\ref{fig:merged_moon_3h} after each system has been evolved for an additional 45 h (48 h in total).}
    \label{fig:merged_moon_48h_1p0}
\end{figure*}

\begin{figure*}[!h]
    \resizebox{\hsize}{!}{\includegraphics{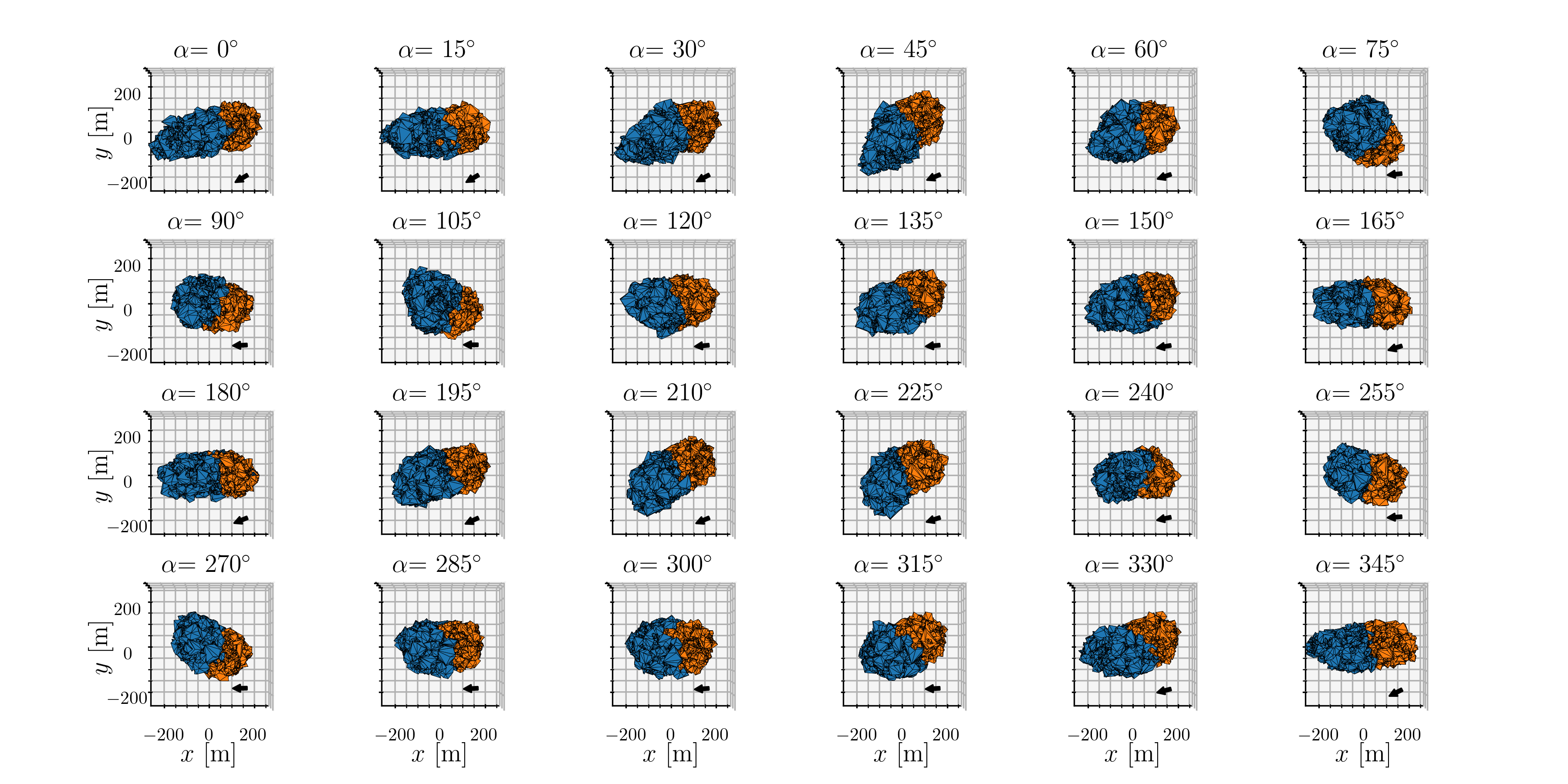}}
    \caption{Same as Fig.~\ref{fig:merged_moon_3h} but after putting each body on a circular orbit with a semi-major axis of $1.2r_0$ and evolving each system for and additional 45 h (48 h in total), where $r_0$ is the distance from the primary at 3 h.}
    \label{fig:merged_moon_48h_1p2}
\end{figure*}

First, we show the resulting shapes of the merged moons from Fig.~\ref{fig:merged_moon_3h} after 45 h of additional simulation (48 h in total) in Fig.~\ref{fig:merged_moon_48h_1p0}. Note that most of the bodies have suffered catastrophic tidal disruption, with $\alpha=300^\circ$ being in the middle of its disruption at the end of the simulation. We also include a plot depicting the corresponding shapes for a case where they were evolved at $1.2r_0$ after 48 h of total simulation in Fig.~\ref{fig:merged_moon_48h_1p2}.  

\begin{figure}[!h]
    \centering
    \resizebox{0.8\hsize}{!}{\includegraphics{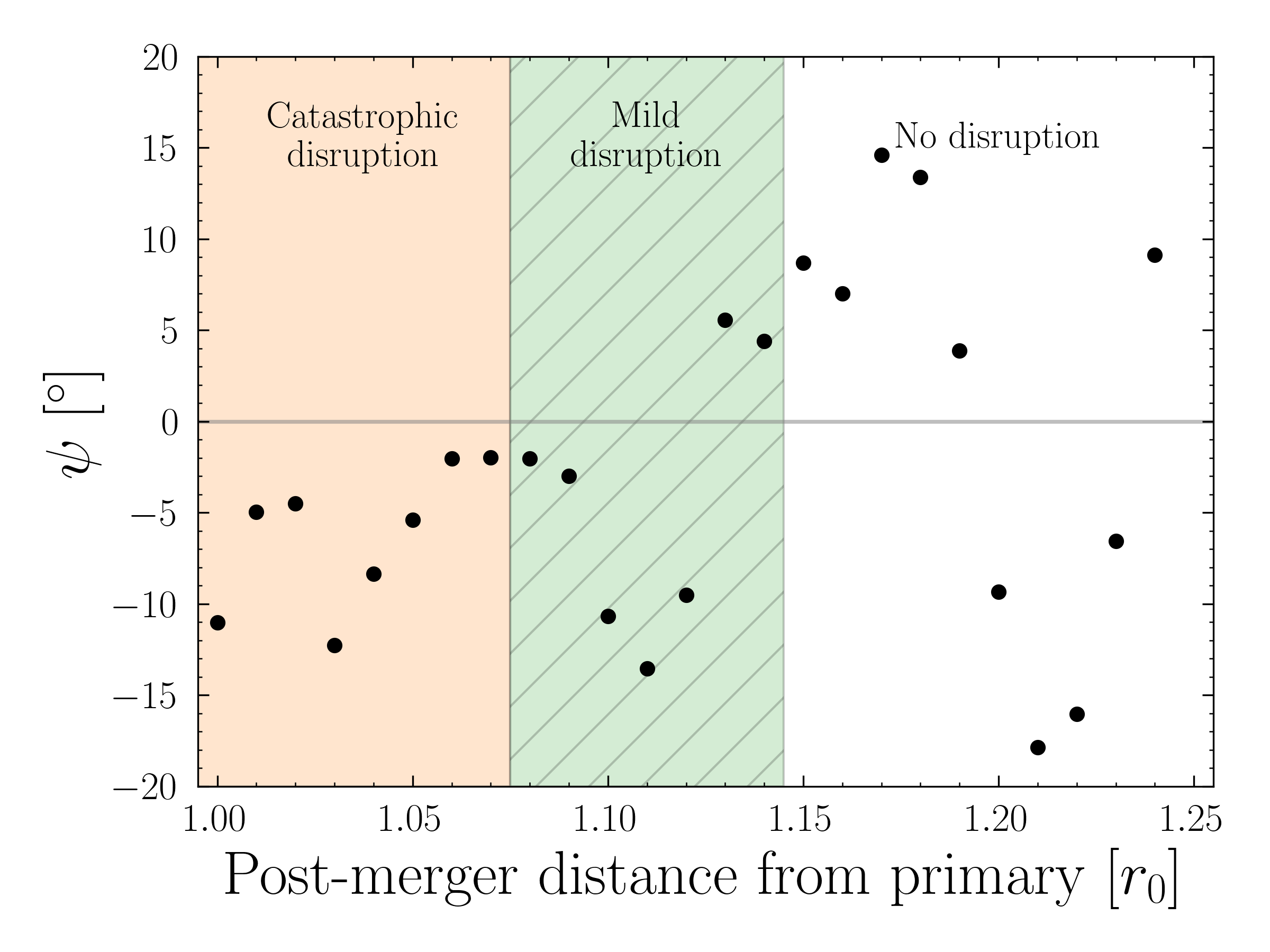}}
    \caption{Yaw angle, $\psi$, at the first time step after mass loss for each case in Fig.~\ref{fig:ab_vals_norm_dist}. Note that the no-disruption cases show the yaw angle at the final time step.}
    \label{fig:yaw_angle_at_mass_loss_norm_dist}
\end{figure}

The next plot in Fig.~\ref{fig:yaw_angle_at_mass_loss_norm_dist} shows the yaw angle for the distance dependency study in Sect.~\ref{section:discussion_tidal_study} when a body first undergoes mass-loss. A majority of the bodies with synchronous rotation in Fig.~\ref{fig:periods_norm_dist} were also tidally locked when the disruption event occurred. For the bodies with $r>1.14r_0$, we simply plotted the yaw angle at the end of the simulation, which explains the variance in rotational period for these objects.

\begin{sidewaystable*}[!ht]
\section{Moon property tables}
    \caption{Physical and orbital properties for the moons in Fig.~\ref{fig:merged_moon_48h_1p0}.}
    \label{tab:moon_properties_48h_1p0}
    \begin{tabular}{l l l l l l l l l l l l l l l l l}
         \hline
         $\alpha\ [^\circ]$ & $m_\mathrm{moon}\ [M_0]$ & $a_0/b_0$ & $b_0/c_0$ & $a_0\ [R_\mathrm{primary}]$ & $\phi_0$ & $a/b$ & $b/c$ & $a\ [R_\mathrm{primary}]$ & $\phi$ & $\gamma\ [^\circ]$ & Class & $a_\mathrm{orb}\ [R_\mathrm{primary}]$ & $e_\mathrm{orb}$ & $i_\mathrm{orb}\ [^\circ]$ & $P_\mathrm{orb}$ [h] & $P_\mathrm{rot}$ [h] \\
         \hline
        0 & 0.36 & 2.45 & 0.97 & 0.51 & 0.36 & 1.45 & 0.72 & 0.22 & 0.20 & - & PS & 6.19 & 0.52 & 1.41 & 47.68 & 11.28\\ 
        15 & 0.35 & 2.21 & 1.10 & 0.50 & 0.35 & 1.14 & 1.29 & 0.22 & 0.27 & - & OS & 6.43 & 0.54 & 1.25 & 50.43 & 8.64\\ 
        30 & 0.34 & 1.97 & 1.18 & 0.47 & 0.34 & 1.25 & 0.72 & 0.19 & 0.27 & - & OS & 6.35 & 0.53 & 1.38 & 49.45 & 8.56\\ 
        45 & 0.34 & 1.82 & 1.25 & 0.46 & 0.34 & 1.37 & 0.67 & 0.20 & 0.24 & - & PS & 5.94 & 0.52 & 1.55 & 44.75 & 8.63\\ 
        60 & 0.33 & 1.63 & 1.29 & 0.43 & 0.35 & 1.48 & 0.75 & 0.22 & 0.24 & - & OS & 5.53 & 0.50 & 1.62 & 40.20 & 8.18\\ 
        75 & 0.90 & 1.43 & 1.41 & 0.41 & 0.33 & 2.06 & 0.88 & 0.41 & 0.31 & - & PS & 2.54 & 0.04 & 1.05 & 12.56 & 15.15\\ 
        90 & 1.00 & 1.92 & 0.68 & 0.39 & 0.33 & 1.66 & 1.23 & 0.41 & 0.30 & - & PS & 2.36 & 0.07 & 0.85 & 11.23 & 11.01\\ 
        105 & 1.00 & 2.05 & 0.69 & 0.40 & 0.33 & 1.88 & 1.29 & 0.45 & 0.27 & 37.79 & OAB & 2.37 & 0.06 & 0.99 & 11.32 & 15.13\\ 
        120 & 0.40 & 1.55 & 1.40 & 0.42 & 0.35 & 1.18 & 0.66 & 0.19 & 0.27 & - & PS & 4.73 & 0.31 & 0.45 & 31.80 & 14.14\\ 
        135 & 0.47 & 1.83 & 1.30 & 0.46 & 0.29 & 1.54 & 1.20 & 0.30 & 0.26 & - & PS & 4.76 & 0.38 & 1.11 & 32.13 & 48.87\\ 
        150 & 0.49 & 2.40 & 0.78 & 0.46 & 0.34 & 1.87 & 0.87 & 0.31 & 0.30 & - & PS & 4.52 & 0.37 & 1.21 & 29.73 & 28.85\\ 
        165 & 0.48 & 2.41 & 0.81 & 0.47 & 0.35 & 1.66 & 1.08 & 0.31 & 0.25 & - & PS & 4.54 & 0.35 & 1.16 & 29.96 & 24.36\\ 
        180 & 0.37 & 2.50 & 0.83 & 0.49 & 0.33 & 1.35 & 0.83 & 0.22 & 0.25 & - & OS & 5.41 & 0.44 & 0.95 & 38.99 & 12.78\\ 
        195 & 0.36 & 2.36 & 0.89 & 0.48 & 0.34 & 1.20 & 1.11 & 0.22 & 0.27 & - & PS & 5.79 & 0.50 & 1.39 & 43.08 & 8.61\\ 
        210 & 0.34 & 2.25 & 0.89 & 0.47 & 0.34 & 1.20 & 0.76 & 0.19 & 0.23 & - & OS & 6.37 & 0.53 & 1.47 & 49.71 & 8.49\\ 
        225 & 0.37 & 1.90 & 1.13 & 0.46 & 0.33 & 1.28 & 1.14 & 0.24 & 0.28 & - & PS & 5.64 & 0.50 & 1.76 & 41.41 & 8.25\\ 
        240 & 0.37 & 1.50 & 1.35 & 0.41 & 0.31 & 1.43 & 0.86 & 0.23 & 0.26 & - & PS & 5.39 & 0.43 & 0.49 & 38.67 & 15.77\\ 
        255 & 1.00 & 1.38 & 1.43 & 0.40 & 0.32 & 1.48 & 1.37 & 0.39 & 0.28 & - & PS & 2.36 & 0.06 & 0.79 & 11.20 & 10.16\\ 
        270 & 1.00 & 1.35 & 1.42 & 0.39 & 0.34 & 2.02 & 0.77 & 0.40 & 0.27 & - & PS & 2.37 & 0.07 & 0.82 & 11.26 & 11.92\\ 
        285 & 1.00 & 1.40 & 1.38 & 0.39 & 0.34 & 1.75 & 1.31 & 0.43 & 0.29 & - & PS & 2.37 & 0.06 & 0.72 & 11.29 & 13.87\\ 
        300 & 1.00 & 1.49 & 1.34 & 0.41 & 0.34 & 4.57 & 1.16 & 0.87 & 0.38 & - & - & 2.34 & 0.04 & 0.98 & 11.10 & 16.09\\ 
        315 & 0.49 & 1.67 & 1.28 & 0.44 & 0.29 & 1.59 & 0.91 & 0.28 & 0.20 & - & PS & 4.28 & 0.33 & 1.57 & 27.43 & 22.08\\ 
        330 & 0.49 & 1.96 & 1.18 & 0.47 & 0.35 & 1.64 & 0.90 & 0.29 & 0.21 & - & PS & 4.50 & 0.35 & 1.31 & 29.53 & 31.58\\ 
        345 & 0.38 & 2.53 & 0.92 & 0.52 & 0.36 & 1.11 & 1.29 & 0.23 & 0.31 & - & OS & 5.78 & 0.50 & 1.44 & 42.95 & 11.61\\  
    \end{tabular}
    \tablefoot{The mass and particle number, as well as geometrical and orbital properties for each case of $\alpha$ in Fig.~\ref{fig:merged_moon_48h_1p0}, where each body has been evolved for 48 h in total. The principal axes are given by $a$, $b$, $c$ and the porosity is given by $\phi$. Subscripts of 0 indicate the reference initial values at 3 h, while the subscript ``orb'' indicates an orbital element. For the classification abbreviations, see the text.}
\end{sidewaystable*}

\begin{sidewaystable*}[!ht]
    \caption{Physical and orbital properties for the moons in Fig.~\ref{fig:merged_moon_48h_1p1}.}
    \label{tab:moon_properties_48h_1p1}
    \begin{tabular}{l l l l l l l l l l l l l l l l l}
         \hline
         $\alpha\ [^\circ]$ & $m_\mathrm{moon}\ [M_0]$ & $a_0/b_0$ & $b_0/c_0$ & $a_0\ [R_\mathrm{primary}]$ & $\phi_0$ & $a/b$ & $b/c$ & $a\ [R_\mathrm{primary}]$ & $\phi$ & $\gamma\ [^\circ]$ & Class & $a_\mathrm{orb}\ [R_\mathrm{primary}]$ & $e_\mathrm{orb}$ & $i_\mathrm{orb}\ [^\circ]$ & $P_\mathrm{orb}$ [h] & $P_\mathrm{rot}$ [h] \\
         \hline
        0 & 0.90 & 2.44 & 0.97 & 0.51 & 0.36 & 2.21 & 0.98 & 0.44 & 0.25 & $-1.2$ & LAB & 2.96 & 0.08 & 0.67 & 15.77 & 16.04\\ 
        15 & 0.35 & 2.21 & 1.11 & 0.50 & 0.35 & 1.33 & 0.82 & 0.21 & 0.27 & - & OS & 6.91 & 0.54 & 1.43 & 56.21 & 17.12\\ 
        30 & 0.91 & 1.96 & 1.20 & 0.47 & 0.34 & 1.93 & 1.18 & 0.43 & 0.32 & $-33.3$ & OAB & 2.93 & 0.06 & 0.50 & 15.50 & 17.21\\ 
        45 & 0.99 & 1.83 & 1.26 & 0.46 & 0.34 & 2.64 & 0.84 & 0.49 & 0.31 & $15.4$ & OAB & 2.61 & 0.05 & 0.48 & 13.05 & 17.25\\ 
        60 & 1.00 & 1.63 & 1.30 & 0.43 & 0.34 & 1.96 & 1.19 & 0.45 & 0.21 & $17.3$ & OAB & 2.60 & 0.07 & 0.63 & 12.97 & 12.54\\ 
        75 & 1.00 & 1.41 & 1.42 & 0.41 & 0.34 & 1.56 & 1.21 & 0.39 & 0.29 & - & PS & 2.64 & 0.07 & 0.56 & 13.24 & 14.56\\ 
        90 & 1.00 & 1.94 & 0.68 & 0.39 & 0.33 & 1.68 & 0.82 & 0.36 & 0.27 & - & PS & 2.64 & 0.03 & 0.60 & 13.30 & 15.19\\ 
        105 & 1.00 & 2.04 & 0.69 & 0.40 & 0.32 & 1.48 & 1.30 & 0.39 & 0.26 & - & PS & 2.65 & 0.06 & 0.53 & 13.38 & 15.49\\ 
        120 & 1.00 & 1.55 & 1.41 & 0.42 & 0.32 & 2.08 & 0.79 & 0.41 & 0.31 & - & PS & 2.66 & 0.05 & 0.58 & 13.43 & 21.45\\ 
        135 & 1.00 & 1.83 & 1.30 & 0.46 & 0.30 & 2.24 & 0.85 & 0.44 & 0.25 & $-18.7$ & OAB & 2.64 & 0.06 & 0.54 & 13.26 & 12.27\\ 
        150 & 1.00 & 2.39 & 0.78 & 0.46 & 0.33 & 2.26 & 0.85 & 0.44 & 0.27 & $-8.9$ & LAB & 2.63 & 0.05 & 0.53 & 13.20 & 12.78\\ 
        165 & 1.00 & 2.41 & 0.81 & 0.47 & 0.34 & 2.13 & 1.18 & 0.47 & 0.29 & $4.7$ & LAB & 2.62 & 0.04 & 0.41 & 13.09 & 18.79\\ 
        180 & 1.00 & 2.49 & 0.83 & 0.48 & 0.33 & 2.48 & 1.19 & 0.53 & 0.29 & $6.4$ & LAB & 2.57 & 0.04 & 0.53 & 12.78 & 15.23\\ 
        195 & 0.45 & 2.38 & 0.89 & 0.48 & 0.33 & 1.44 & 1.15 & 0.28 & 0.29 & - & PS & 4.62 & 0.33 & 0.25 & 30.71 & 20.48\\ 
        210 & 0.47 & 2.29 & 0.88 & 0.48 & 0.33 & 1.61 & 1.21 & 0.31 & 0.24 & - & PS & 4.72 & 0.34 & 0.47 & 31.73 & 52.59\\ 
        225 & 1.00 & 1.90 & 1.13 & 0.46 & 0.33 & 2.97 & 1.10 & 0.58 & 0.28 & $0.5$ & LAB & 2.54 & 0.04 & 0.52 & 12.56 & 14.48\\ 
        240 & 1.00 & 1.49 & 1.36 & 0.41 & 0.31 & 1.50 & 1.25 & 0.39 & 0.29 & $-61.2$ & SAB & 2.63 & 0.07 & 0.50 & 13.22 & 15.04\\ 
        255 & 1.00 & 1.39 & 1.44 & 0.40 & 0.32 & 1.79 & 0.77 & 0.37 & 0.28 & $-25.6$ & OAB & 2.64 & 0.04 & 0.51 & 13.28 & 17.24\\ 
        270 & 1.00 & 1.36 & 1.42 & 0.39 & 0.34 & 1.37 & 1.27 & 0.36 & 0.30 & - & PS & 2.64 & 0.03 & 0.46 & 13.28 & 12.48\\ 
        285 & 1.00 & 1.40 & 1.37 & 0.39 & 0.35 & 1.49 & 1.25 & 0.38 & 0.29 & - & PS & 2.66 & 0.07 & 0.50 & 13.41 & 17.48\\ 
        300 & 1.00 & 1.49 & 1.34 & 0.41 & 0.32 & 1.56 & 1.23 & 0.39 & 0.20 & $-73.0$ & SAB & 2.66 & 0.05 & 0.59 & 13.46 & 19.39\\ 
        315 & 1.00 & 1.68 & 1.28 & 0.44 & 0.31 & 1.81 & 1.15 & 0.43 & 0.32 & $-5.9$ & LAB & 2.64 & 0.05 & 0.60 & 13.29 & 12.88\\ 
        330 & 0.89 & 1.95 & 1.19 & 0.47 & 0.32 & 1.76 & 1.12 & 0.39 & 0.23 & $-13.1$ & OAB & 2.92 & 0.05 & 0.59 & 15.47 & 7.74\\ 
        345 & 0.89 & 2.53 & 0.92 & 0.52 & 0.36 & 2.05 & 1.06 & 0.43 & 0.29 & - & PS & 2.95 & 0.08 & 0.70 & 15.72 & 17.47\\ 
    \end{tabular}
    \tablefoot{The same as Table~\ref{tab:moon_properties_48h_1p0} for the case where each merged moon from Fig.~\ref{fig:merged_moon_3h} has been pushed out to a distance of $1.1r_0$ before letting it evolve for 45 h. The small deviations in e.g.~$a_0/b_0$ to the corresponding values in Table~\ref{tab:moon_properties_48h_1p0} are due to reductions in numerical precision when writing and reading data.}
\end{sidewaystable*}

\begin{sidewaystable*}[!ht]
    \caption{Physical and orbital properties for the moons in Fig.~\ref{fig:merged_moon_48h_1p2}.}
    \label{tab:moon_properties_48h_1p2}
    \begin{tabular}{l l l l l l l l l l l l l l l l l}
         \hline
         $\alpha\ [^\circ]$ & $m_\mathrm{moon}\ [M_0]$ & $a_0/b_0$ & $b_0/c_0$ & $a_0\ [R_\mathrm{primary}]$ & $\phi_0$ & $a/b$ & $b/c$ & $a\ [R_\mathrm{primary}]$ & $\phi$ & $\gamma\ [^\circ]$  & Class & $a_\mathrm{orb}\ [R_\mathrm{primary}]$ & $e_\mathrm{orb}$ & $i_\mathrm{orb}\ [^\circ]$ & $P_\mathrm{orb}$ [h] & $P_\mathrm{rot}$ [h] \\
         \hline
     0 & 1.00 & 2.44 & 0.97 & 0.51 & 0.36 & 2.33 & 1.04 & 0.49 & 0.25 & $-3.44$ & LAB & 2.89 & 0.04 & 0.60 & 15.22 & 11.77\\ 
    15 & 1.00 & 2.21 & 1.11 & 0.50 & 0.35 & 2.16 & 1.06 & 0.46 & 0.26 & - & PS & 2.90 & 0.03 & 0.62 & 15.26 & 14.55\\ 
    30 & 1.00 & 1.96 & 1.20 & 0.47 & 0.34 & 2.00 & 1.15 & 0.45 & 0.31 & $12.81$ & OAB & 2.91 & 0.02 & 0.65 & 15.36 & 30.13\\ 
    45 & 1.00 & 1.83 & 1.26 & 0.46 & 0.34 & 1.84 & 1.19 & 0.44 & 0.30 & $29.32$ & OAB & 2.89 & 0.03 & 0.64 & 15.22 & 16.69\\ 
    60 & 1.00 & 1.63 & 1.30 & 0.43 & 0.34 & 1.70 & 1.16 & 0.41 & 0.31 & - & PS & 2.88 & 0.04 & 0.48 & 15.13 & 11.67\\ 
    75 & 1.00 & 1.41 & 1.42 & 0.41 & 0.34 & 1.35 & 1.23 & 0.36 & 0.29 & - & PS & 2.87 & 0.03 & 0.59 & 15.06 & 13.46\\ 
    90 & 1.00 & 1.94 & 0.68 & 0.39 & 0.33 & 1.62 & 0.80 & 0.35 & 0.26 & - & PS & 2.89 & 0.04 & 0.53 & 15.17 & 16.88\\ 
    105 & 1.00 & 2.04 & 0.69 & 0.40 & 0.32 & 1.36 & 1.28 & 0.36 & 0.29 & - & PS & 2.87 & 0.02 & 0.60 & 15.05 & 11.92\\ 
    120 & 1.00 & 1.55 & 1.41 & 0.42 & 0.32 & 1.56 & 1.24 & 0.39 & 0.29 & $50.11$ & SAB & 2.89 & 0.05 & 0.42 & 15.18 & 12.10\\ 
    135 & 1.00 & 1.83 & 1.30 & 0.46 & 0.31 & 1.72 & 1.15 & 0.41 & 0.26 & $-44.01$ & OAB & 2.90 & 0.03 & 0.53 & 15.30 & 17.51\\ 
    150 & 1.00 & 2.39 & 0.78 & 0.46 & 0.33 & 1.73 & 1.18 & 0.41 & 0.30 & - & PS & 2.90 & 0.03 & 0.53 & 15.31 & 17.49\\ 
    165 & 1.00 & 2.41 & 0.81 & 0.47 & 0.34 & 1.86 & 1.15 & 0.43 & 0.27 & $-27.12$ & OAB & 2.91 & 0.02 & 0.58 & 15.36 & 27.57\\ 
    180 & 1.00 & 2.49 & 0.83 & 0.48 & 0.33 & 1.93 & 1.16 & 0.44 & 0.26 & - & PS & 2.91 & 0.02 & 0.57 & 15.36 & 28.67\\ 
    195 & 1.00 & 2.38 & 0.89 & 0.48 & 0.33 & 2.24 & 0.86 & 0.44 & 0.28 & $28.51$ & OAB & 2.91 & 0.02 & 0.52 & 15.38 & 33.38\\ 
    210 & 1.00 & 2.29 & 0.88 & 0.48 & 0.33 & 1.89 & 1.14 & 0.44 & 0.29 & - & PS & 2.90 & 0.02 & 0.52 & 15.32 & 26.61\\ 
    225 & 1.00 & 1.90 & 1.13 & 0.46 & 0.33 & 1.74 & 1.06 & 0.41 & 0.30 & $40.52$ & OAB & 2.88 & 0.05 & 0.52 & 15.13 & 11.98\\ 
    240 & 1.00 & 1.49 & 1.36 & 0.41 & 0.31 & 1.40 & 1.20 & 0.36 & 0.30 & $67.16$ & SAB & 2.88 & 0.01 & 0.41 & 15.09 & 13.20\\ 
    255 & 1.00 & 1.39 & 1.44 & 0.40 & 0.32 & 1.68 & 0.78 & 0.35 & 0.30 & $52.61$ & SAB & 2.88 & 0.03 & 0.41 & 15.11 & 13.52\\ 
    270 & 1.00 & 1.36 & 1.42 & 0.39 & 0.34 & 1.62 & 0.80 & 0.35 & 0.31 & - & PS & 2.88 & 0.04 & 0.41 & 15.13 & 14.96\\ 
    285 & 1.00 & 1.40 & 1.37 & 0.39 & 0.35 & 1.41 & 1.24 & 0.37 & 0.26 & - & PS & 2.87 & 0.02 & 0.43 & 15.04 & 10.72\\ 
    300 & 1.00 & 1.49 & 1.34 & 0.41 & 0.32 & 1.77 & 0.83 & 0.37 & 0.26 & - & PS & 2.89 & 0.05 & 0.40 & 15.17 & 12.82\\ 
    315 & 1.00 & 1.68 & 1.28 & 0.44 & 0.31 & 1.83 & 0.87 & 0.39 & 0.30 & $83.60$ & SAB & 2.88 & 0.06 & 0.47 & 15.15 & 11.58\\ 
    330 & 1.00 & 1.95 & 1.19 & 0.47 & 0.32 & 1.92 & 1.12 & 0.44 & 0.25 & $-36.80$ & OAB & 2.91 & 0.02 & 0.58 & 15.39 & 31.46\\ 
    345 & 1.00 & 2.53 & 0.92 & 0.52 & 0.36 & 2.26 & 1.06 & 0.48 & 0.26 & - & PS & 2.90 & 0.04 & 0.60 & 15.26 & 12.55\\ 
    \end{tabular}
    \tablefoot{The same as Table~\ref{tab:moon_properties_48h_1p0} for the case where each merged moon from Fig.~\ref{fig:merged_moon_3h} has been pushed out to a distance of $1.2r_0$ before letting it evolve for 45 h.}
\end{sidewaystable*}

\begin{sidewaystable*}[!ht]
    \caption{Physical and orbital properties for the moons in Fig.~\ref{fig:merged_moon_48h_norm_dist}.}
    \label{tab:moon_properties_48h_norm_dist}
    \begin{tabular}{l l l l l l l l l l l l l l l l l}
         \hline
         $r\ [r_0]$ & $m_\mathrm{moon}\ [M_0]$ & $a_0/b_0$ & $b_0/c_0$ & $a_0\ [R_\mathrm{primary}]$ & $\phi_0$ & $a/b$ & $b/c$ & $a\ [R_\mathrm{primary}]$ & $\phi$ & $\gamma\ [^\circ]$ & Class & $a_\mathrm{orb}\ [R_\mathrm{primary}]$ & $e_\mathrm{orb}$ & $i_\mathrm{orb}\ [^\circ]$ & $P_\mathrm{orb}$ [h] & $P_\mathrm{rot}$ [h] \\
         \hline
        1.00 & 0.35 & 2.44 & 0.97 & 0.51 & 0.36 & 1.48 & 0.78 & 0.23 & 0.31 & - & OS & 6.31 & 0.53 & 1.22 & 49.06 & 8.28\\ 
        1.01 & 0.37 & 2.44 & 0.97 & 0.51 & 0.36 & 1.09 & 1.26 & 0.22 & 0.27 & - & OS & 5.09 & 0.44 & 0.81 & 35.55 & 12.30\\ 
        1.02 & 0.35 & 2.44 & 0.97 & 0.51 & 0.36 & 1.43 & 0.78 & 0.22 & 0.30 & - & OS & 4.92 & 0.41 & 0.76 & 33.74 & 9.73\\ 
        1.03 & 0.37 & 2.44 & 0.97 & 0.51 & 0.36 & 1.36 & 0.80 & 0.22 & 0.32 & - & OS & 4.89 & 0.41 & 0.81 & 33.44 & 10.97\\ 
        1.04 & 0.37 & 2.44 & 0.97 & 0.51 & 0.36 & 1.19 & 1.24 & 0.23 & 0.31 & - & OS & 4.89 & 0.40 & 0.95 & 33.42 & 16.50\\ 
        1.05 & 0.36 & 2.44 & 0.97 & 0.51 & 0.36 & 1.08 & 1.26 & 0.22 & 0.23 & - & OS & 5.33 & 0.42 & 1.08 & 38.04 & 15.49\\ 
        1.06 & 0.35 & 2.44 & 0.97 & 0.51 & 0.36 & 1.08 & 1.27 & 0.21 & 0.29 & - & OS & 5.56 & 0.44 & 1.18 & 40.61 & 17.05\\ 
        1.07 & 0.35 & 2.44 & 0.97 & 0.51 & 0.36 & 1.39 & 0.83 & 0.22 & 0.31 & - & OS & 5.23 & 0.37 & 1.81 & 36.97 & 16.74\\ 
        1.08 & 0.89 & 2.44 & 0.97 & 0.51 & 0.36 & 2.27 & 0.98 & 0.45 & 0.31 & $-5.12$ & LAB & 2.91 & 0.07 & 0.59 & 15.37 & 16.52\\ 
        1.09 & 0.89 & 2.44 & 0.97 & 0.51 & 0.36 & 2.26 & 0.97 & 0.45 & 0.28 & $-0.37$ & LAB & 2.92 & 0.08 & 0.63 & 15.43 & 15.00\\ 
        1.10 & 0.90 & 2.44 & 0.97 & 0.51 & 0.36 & 2.21 & 0.98 & 0.44 & 0.25 & $-1.16$ & LAB & 2.96 & 0.08 & 0.67 & 15.77 & 16.04\\ 
        1.11 & 0.89 & 2.44 & 0.97 & 0.51 & 0.36 & 2.12 & 0.98 & 0.43 & 0.26 & - & PS & 2.98 & 0.07 & 0.64 & 15.93 & 19.82\\ 
        1.12 & 0.89 & 2.44 & 0.97 & 0.51 & 0.36 & 2.22 & 0.96 & 0.44 & 0.27 & - & PS & 3.01 & 0.09 & 0.74 & 16.14 & 13.91\\ 
        1.13 & 0.90 & 2.44 & 0.97 & 0.51 & 0.36 & 2.23 & 0.97 & 0.44 & 0.27 & $-0.56$ & LAB & 2.99 & 0.04 & 0.60 & 15.98 & 11.62\\ 
        1.14 & 0.90 & 2.44 & 0.97 & 0.51 & 0.36 & 2.56 & 1.01 & 0.49 & 0.30 & - & PS & 3.02 & 0.07 & 0.54 & 16.21 & 15.03\\ 
        1.15 & 1.00 & 2.44 & 0.97 & 0.51 & 0.36 & 2.82 & 0.98 & 0.54 & 0.20 & - & PS & 2.81 & 0.02 & 0.48 & 14.56 & 18.60\\ 
        1.16 & 1.00 & 2.44 & 0.97 & 0.51 & 0.36 & 2.73 & 0.97 & 0.52 & 0.23 & - & PS & 2.84 & 0.02 & 0.50 & 14.80 & 19.85\\ 
        1.17 & 1.00 & 2.44 & 0.97 & 0.51 & 0.36 & 2.66 & 0.96 & 0.51 & 0.20 & $-4.01$ & LAB & 2.86 & 0.02 & 0.53 & 14.96 & 15.94\\ 
        1.18 & 1.00 & 2.44 & 0.97 & 0.51 & 0.36 & 2.62 & 0.96 & 0.51 & 0.24 & $-2.89$ & LAB & 2.87 & 0.03 & 0.55 & 15.05 & 12.92\\ 
        1.19 & 1.00 & 2.44 & 0.97 & 0.51 & 0.36 & 2.51 & 0.97 & 0.50 & 0.32 & $-1.49$ & LAB & 2.88 & 0.03 & 0.57 & 15.13 & 11.46\\ 
        1.20 & 1.00 & 2.44 & 0.97 & 0.51 & 0.36 & 2.33 & 1.04 & 0.49 & 0.25 & $-3.44$ & LAB & 2.89 & 0.04 & 0.60 & 15.22 & 11.77\\ 
        1.21 & 1.00 & 2.44 & 0.97 & 0.51 & 0.36 & 2.37 & 0.95 & 0.48 & 0.27 & $-1.77$ & LAB & 2.91 & 0.03 & 0.62 & 15.34 & 13.76\\ 
        1.22 & 1.00 & 2.44 & 0.97 & 0.51 & 0.36 & 2.32 & 0.94 & 0.47 & 0.28 & $-1.94$ & LAB & 2.92 & 0.03 & 0.64 & 15.44 & 17.94\\ 
        1.23 & 1.00 & 2.44 & 0.97 & 0.51 & 0.36 & 2.26 & 0.94 & 0.46 & 0.26 & $-2.90$ & LAB & 2.93 & 0.03 & 0.63 & 15.53 & 24.84\\ 
        1.24 & 1.00 & 2.44 & 0.97 & 0.51 & 0.36 & 2.18 & 0.95 & 0.45 & 0.24 & $-5.19$ & LAB & 2.94 & 0.03 & 0.59 & 15.58 & 32.17\\ 
        1.25 & 1.00 & 2.44 & 0.97 & 0.51 & 0.36 & 2.07 & 1.06 & 0.45 & 0.26 & $-7.58$ & LAB & 2.94 & 0.03 & 0.57 & 15.62 & 30.17\\    
    \end{tabular}
    \tablefoot{Similar to Table~\ref{tab:moon_properties_48h_1p0} but shows post-merger properties of the largest remnant for varying distance shifts\tablefootmark{a} for the case of $\alpha = 0^\circ$ in Fig.~\ref{fig:merged_moon_48h_norm_dist}. The DEEVE principal axis data and porosity values at 48 h can be compared to their initial counterparts at 3 h, indicated by the subscript 0.\\
    \tablefoottext{a}{We note here that the dynamical properties of $1.00r_0$ are different from the corresponding case in Table~\ref{tab:moon_properties_48h_1p0} due to the merged moon being given a different post-merger velocity corresponding to a circular orbit in this scenario, for consistency with the remaining $kr_0$ cases.}}
\end{sidewaystable*}

\end{appendix}

\end{document}